\documentclass[12pt]{article}
\usepackage{graphicx}
\usepackage{natbib}
\usepackage{multirow}
\usepackage{dcolumn,booktabs}
\usepackage{algorithm,algorithmicx,algpseudocode,algcompatible}
\usepackage{authblk} % Manages authors' affiliations
\usepackage{url} 
\usepackage[table]{xcolor}

\usepackage[font=small]{caption} % To reduce caption font size
\usepackage{amsmath,amssymb,bm,amsthm} % amsfonts,mathdots

\graphicspath{{Figures}}

\DeclareMathOperator*{\argmax}{argmax}

\newcommand{\bhat}{\bm{\hat}}
\newcommand{\btilde}{\bm{\tilde}}
\newcommand{\bbar}{\bm{\bar}}
\newcommand{\cond}{\boldsymbol{|}}

\newcommand{\cG}{\mathcal{G}}
\newcommand{\cW}{\mathcal{W}}
\newcommand{\cX}{\mathcal{X}}
\newcommand{\cY}{\mathcal{Y}}
\newcommand{\cZ}{\mathcal{Z}}
\newcommand{\PAC}{\mbox{PAC}}

\newcommand{\halpha}{\widehat{\alpha}}
\newcommand{\hmu}{\widehat{\mu}}
\newcommand{\hpi}{\widehat{\pi}}

\newcommand{\bzero}{\boldsymbol{0}}
\newcommand{\bone}{\boldsymbol{1}}

\newcommand{\tp}{\btilde p}
\newcommand{\hp}{\bhat p}

\newcommand{\bv}{\boldsymbol v}
\newcommand{\bw}{\boldsymbol w}
\newcommand{\bx}{\boldsymbol x}

\newcommand{\bW}{\boldsymbol W}

\newcommand{\bmu}{\boldsymbol \mu}
\newcommand{\bhmu}{\boldsymbol{\widehat{\mu}}}

\newcommand{\bSigma}{\boldsymbol \Sigma}
\newcommand{\bhSigma}{\boldsymbol{\widehat{\Sigma}}}

\newcolumntype{M}[1]{>{\centering\arraybackslash}m{#1}}

\definecolor{orange1}{RGB}{255,128,0}
\definecolor{purple2}{RGB}{102,0,204}
\definecolor{blue}{RGB}{0,0,255}
\definecolor{red}{RGB}{255,0,0}

\newcommand{\N}{\mathcal N}
\newcommand{\Bern}{\mathrm{Bernoulli}}
\newcommand{\Lap}{\mathrm{Laplace}}

\newcommand{\MD}{\mathrm{MD}}

\begin{document}

\def\spacingset#1{\renewcommand{\baselinestretch}
{#1}\small\normalsize} \spacingset{1}

%============================================================

\title{\bf Cellwise Robust Discriminant Analysis}
\author[1]{Fabio Centofanti}
\author[1]{Can Hakan Da\u{g}{\i}d{\i}r}
\author[1]{Mia Hubert}
\author[1]{\mbox{Peter J. Rousseeuw}}

\affil[1]{Section of Statistics and Data Science,
          Department of Mathematics, KU Leuven, Belgium}

\setcounter{Maxaffil}{0}
\renewcommand\Affilfont{\itshape\small}
\date{May 28, 2026}          
\maketitle

\bigskip
\begin{abstract}
Classical discriminant analysis (DA) is based 
on the mean and \mbox{empirical}
covariance matrix
of each class, both of which are sensitive to
outliers in the data. In the past the focus was
on casewise outliers, that is, datapoints 
that lie far away. But nowadays there is
increasing interest in cellwise outliers, that
are unexpected entries in the data matrix.
Removing an entire case because it has one or a 
few outlying cells would lose much information.
Cellwise robust methods aim to detect the 
outlying cells and to preserve the information 
in the other cells. We propose a DA method
that is trained by estimating the location and
covariance of each class by cellwise and
casewise robust estimators, that can also handle 
NA's. The main novelty of our approach is in 
the prediction on test data, that may contain 
outlying cells and NA's themselves. The
new robust discriminant function is derived 
from a novel statistical model by penalized 
maximum likelihood. We focus on 
quadratic DA, but also cover the setting of 
linear DA. The new cellQDA and cellLDA methods 
perform well in simulation. The approach
is illustrated on real data, and the results 
are interpreted with the help of graphical 
displays.
\end{abstract}

\noindent {\it Keywords:}
Cellwise outliers, Linear discriminant
analysis, Quadratic discriminant analysis,
Robust statistics, Supervised classification.

\newpage
\spacingset{1.5}

%========================================
\section{Introduction} \label{sec:intro}

Discriminant analysis (DA) is a supervised 
classification method for multivariate data. 
Classical Linear Discriminant Analysis (LDA) 
and Classical Quadratic Discriminant Analysis
(QDA) are derived under the assumption of 
multivariate normal classes. In this setting, 
the center of a class is estimated 
by its arithmetic mean, and its scatter matrix
is estimated by its empirical
covariance matrix. However, both of these
estimators are sensitive to outliers. The
presence of outliers can therefore hurt the 
classification accuracy.

The search for robust DA methods has 
traditionally focused on casewise outliers, 
that is, cases that lie far from the 
majority of their class. Substantial
progress has been made in this area by 
inserting robust estimates of the center and 
scatter matrix of each class, as done by 
\cite{He:Discrim}, \cite{Croux:Discrim}, 
\cite{RQDA}, \cite{filzmoser2012}, 
\cite{pinheiro2025}, and
\cite{Becquart2026}.

Casewise robust methods downweight or discard 
entire rows of the data matrix, which is
appropriate when the outlying cases are members
of a different population. But in recent
times people have become aware that there is 
also another common type of contamination, 
the so-called cellwise outliers.
In that setting only a few cells (entries) 
of a case may be suspect, while the 
remaining entries still contain valuable 
information. Even a relatively small
percentage of outlying cells in the data
matrix can pollute over half of the cases.
When that happens, casewise robust estimators
may be ineffective. Moreover, an outlying 
cell may be difficult to detect, as it
does not have to be a marginal outlier: it 
suffices that its value differs markedly from 
what would be expected based on the relations
between the variables. For instance, a
child's weight may be unremarkable by itself, 
but incompatible with its age and height.

Therefore there is a growing need for methods 
to detect outlying cells and/or provide 
robust estimates when they occur. 
In particular, cellwise robust discriminant 
analysis has received limited attention in 
the literature, in spite of the practical
importance of this problem. A notable 
exception is \citet{aerts2017}, who estimated
the scatter matrix of each class by the
cellwise robust method of \cite{croux2016}.
This approach yielded a robustly trained DA 
model. However, it did not yet provide a
robust out-of-sample prediction method. Such 
a mechanism would be very useful when 
dealing with test data that themselves
contain outlying cells.

\begin{figure}[!ht]
\vspace{4mm}
\centering
\includegraphics[width=0.6\linewidth]
  {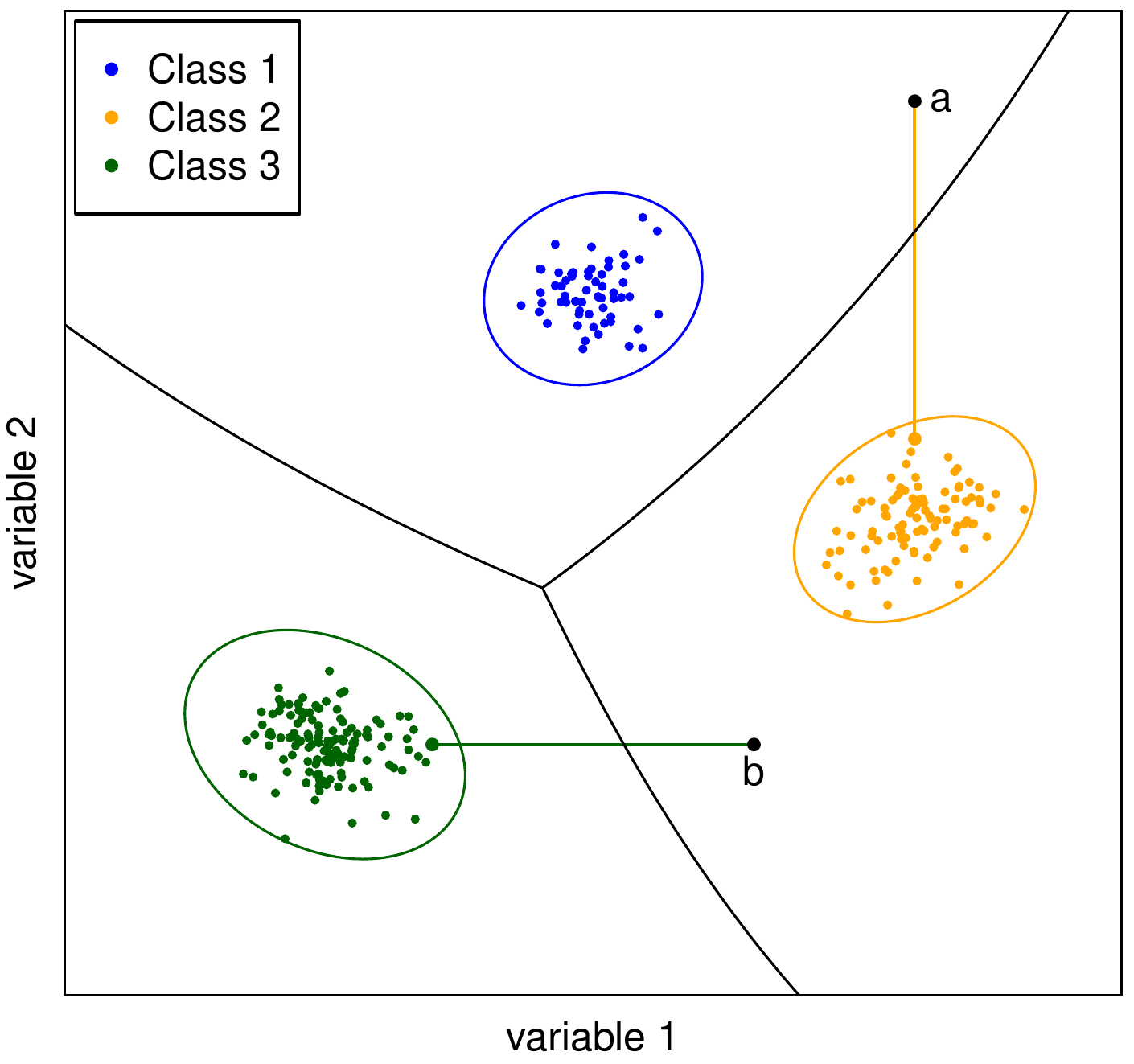}\\
\caption{Illustrative example with 3 classes
and their QDA boundaries. The out-of-sample
points \textbf{a} and \textbf{b} each contain
an outlying cell. The usual discriminant rule 
would misclassify them.}
\label{fig:toy1_a}
\end{figure}

To illustrate this issue, 
Figure~\ref{fig:toy1_a} shows a
toy example with three classes. Suppose that
the training data was clean, so that the
centers and covariance matrices of the
classes are accurate. They yield the 
tolerance ellipses in the plot, and the 
quadratic boundaries of QDA shown as curved 
lines. The points in the figure are
out-of-sample new data. Point \textbf{a}
of class 2 has an outlying cell in its 
second variable so it ends up higher, in the
decision region of class 1. Therefore
the usual discriminant rule would assign it 
to class 1, even though it is far away from
that class as well. Similarly, point 
\textbf{b} of class 3 would be assigned to 
class 2 due to its outlying cell in the
first variable. Our task is to come up 
with an approach to assign points with 
outlying cells to the right class, also in 
higher dimensions where visual inspection 
would not suffice. We will revisit this 
example later.

In this paper we propose a new cellwise 
robust approach to both LDA and QDA, that 
we will call cellLDA and cellQDA. 
Our methodology addresses both stages of 
the classification pipeline. In the 
training phase, we use the recent cellwise 
MCD estimator of \cite{cellMCD} to obtain 
a location and scatter matrix of each class.

The main novelty of our work is in the
second stage, where our out-of-sample 
prediction method can handle outlying 
cells in the test data. For 
this purpose it adopts a
new statistical model that extends the
existing cellwise contamination model.
After fitting this model on the training
data, it is applied out-of-sample by
flagging outlying cells in new cases,
and using a new robust discriminant 
function. It is worth noting that all
the components of this approach, from the
estimation of the centers and the 
covariance matrices in training, to the
flagging of outlying cells out-of-sample,
and the derivation of the new discriminant
function, are all based on the same 
penalized maximum likelihood. Due to
this unification, the approach also 
handles missing values in a natural way. 

To enhance interpretability, we also 
provide a graphical display that enables 
users to visually distinguish between clean 
cases, cases that are classified correctly 
despite some outlying cells, and casewise 
outliers.

The remainder of the paper is organized as 
follows. Section~\ref{sec:method} describes
the proposed methodology for in-sample and
out-of-sample data. Section~\ref{sec:sim} 
evaluates the performance 
of cellLDA and cellQDA through simulation, 
under various contamination scenarios.
Section~\ref{sec:realdata} illustrates the 
approach on a real dataset of nutritional 
values. Section~\ref{sec:conc} concludes.

%================================================
\section{Methodology} \label{sec:method}

Suppose we have training data with
$G$ classes $g = 1,\ldots,G$. Each class
contains $d$-variate points $\bx_{g,i}$ for 
$i=1,\ldots,n_g$\,, and $n=\sum_{g=1}^G n_g$\,. 
The goal of discriminant analysis (DA) is to 
construct a classification rule based on the
training data, that generalizes well to new 
data. 

%================================================
\subsection{Classical discriminant analysis}
\label{sec:DA}

The classical framework assumes that each
class $g$ contains $d$-variate points  
that were generated by a multivariate 
normal distribution. Formally, it assumes
that the random vector $\cX$ follows the 
generative model
\[
  \cX \cond (\cG=g) \sim \N(\bmu_g,\bSigma_g)
  \quad \mbox{with} \quad P(\cG=g)=\pi_g
\]
where $\bmu_g$ and $\bSigma_g$ are the 
class-specific location vector and scatter 
matrix. This is called a normal mixture 
model. The random variable $\cG$ is the 
class membership, so $\pi_g > 0$ is the 
prior probability of class $g$, with 
$\sum_{g=1}^G \pi_g = 1$.

The general form of this model allows each 
class to have its own parameters $\bmu_g$ 
and $\bSigma_g$\,. In this setting the 
discriminant function of class $g$ in a 
$d$-variate case $\bx$ is
\begin{equation} \label{eq:cqda}
 \delta_g^{\mathrm{QDA}}(\bx)=
 -\tfrac12\log|\bSigma_g|
 -\tfrac12(\bx-\bmu_g)^\top\bSigma_g^{-1}
 (\bx-\bmu_g) +\log\pi_g\;.
\end{equation}
The classifier then assigns the case $\bx$ 
to the class $g$ with the highest 
$\delta_g^{\mathrm{QDA}}(\bx)$. This is
called Quadratic Discriminant Analysis (QDA)
because the discriminant function is 
quadratic in $\bx$. It yields curved 
decision boundaries between classes. 

In practice, estimating a separate 
covariance matrix $\bSigma_g$ for each class 
is infeasible when some classes contain 
too few cases. A common approach is then to 
assume a shared covariance matrix $\bSigma$ 
across classes, that is, 
$\bSigma_g = \bSigma$ for all $g$.
Under this homogeneity assumption, the 
discriminant function simplifies to
\begin{equation}\label{eq:clda}
\delta_g^{\mathrm{LDA}}(\bx)
= \bx^\top\bSigma^{-1}\bmu_g
-\tfrac{1}{2}\bmu_g^\top\bSigma^{-1}\bmu_g
+\log\pi_g\,,
\end{equation}
as the terms $-\tfrac12\log|\bSigma_g|
 -\tfrac{1}{2}\bx^\top\bSigma^{-1}\bx$
have dropped out because they do not depend 
\mbox{on $g$}. Now the discriminant function 
is linear in $\bx$, so this method is 
referred to as Linear Discriminant Analysis
(LDA). It yields linear decision 
boundaries.

The classical estimators of 
$(\bmu_g, \bSigma_g)$ are the mean and 
the empirical covariance matrix. 
Both are sensitive to cellwise outliers, 
and so are the discriminant functions 
\eqref{eq:cqda}--\eqref{eq:clda}. We 
therefore need to (i) robustly estimate 
the class parameters in training, and (ii) 
construct a discriminant rule that is robust
against cellwise contamination in new data.

%================================================
\subsection{Robustly estimating class 
  characteristics}  
\label{sec:cellMCD}

In the QDA setting, we will estimate 
the class parameters 
by the cellMCD method of \cite{cellMCD}.
Consider a class $g$ with members 
$\bx_{g,i}$ for $i=1,\ldots,n_g$\,.

On it, cellMCD optimizes an objective 
function to estimate $(\bmu_g, \bSigma_g)$ 
as well as an $n_g \times d$ binary matrix
$\bW_g$\,. The $i$th row $\bw_{g,i}$ of 
$\bW_g$ has $w_{g,ij} = 0$ when the cell 
$x_{g,ij}$ is flagged as outlying or NA, 
and $w_{g,ij} = 1$ indicates that it is 
unflagged. We denote the set of indices 
of the flagged cells in case $i$ as
$m_i := m(\bw_{g,i})=\{j : w_{g,ij} = 0\}$
and that of the unflagged cells as
$o_i := o(\bw_{g,i})=\{j : w_{g,ij} = 1\}.$
With this notation, cellMCD minimizes
\begin{eqnarray} \label{eq:cellMCD}
\sum_{i=1}^{n_g} \Big(\log
  |\bSigma_g^{(o_i)}| + |o_i|\log(2\pi) 
  + \MD^2(\bx_{g,i},\bw_{g,i},\bmu_g,\bSigma_g) 
  \Big) + \sum_{j=1}^d q_{gj} \| 
  \bone_{n_g} - (\bW_g)_{\cdot j} \|_0  \\
  \text{under the constraints } 
  \lambda_{\min}(\bSigma_g) \geqslant 
  a > 0 \text{ and } \nonumber \\
  \|(\bW_g)_{\cdot j}\|_0 \geqslant h_g 
  \text{ for all } j = 1, \dots, d\;.
  \nonumber  
\end{eqnarray}
Here the superscript $\phantom{}^{(o_i)}$ 
restricts a vector or square matrix to 
indices $j$ in $o_i$\,,  the 
number of unflagged cells of $\bx_{g,i}$
is $|o_i|$, and
\[
\MD^2(\bx_{g,i},\bw_{g,i},\bmu_g,\bSigma_g) := 
 (\bx_{g,i}^{(o_i)} - \bmu_g^{(o_i)})^\top 
 (\bSigma_g^{(o_i)})^{-1}
 (\bx_{g,i}^{(o_i)} - \bmu_g^{(o_i)})
\]
is the partial squared Mahalanobis distance.
Note that the first sum is the usual observed
negative log-likelihood for incomplete data.
Moreover, $\bone_{n_g}$ is a column vector 
of $n_g$ ones, and  
$(\bW_g)_{\cdot j}$ is the 
$j$th column of $\bW_g$. Therefore
$\|\bone_{n_g} - (\bW_g)_{\cdot j}\|_0$ 
is the
number of flagged cells in the $j$th 
variable of class $g$. The $q_{gj} > 0$ 
are tuning parameters.
The $\lambda_{\min}(\bSigma_g)$ in 
the first constraint is the smallest 
eigenvalue of $\bSigma_g$\,, so 
$\bSigma_g$ is positive definite. It 
follows that for any square submatrix 
$\bSigma_g^{(o)}$ with nonempty set $o$
and any $\bx^{(o)} \neq \bzero$ we have 
\mbox{$(\bx^{(o)})^\top\bSigma_g^{(o)}
\bx^{(o)} = (\bx^*)^\top\bSigma_g\bx^* 
\geqslant 
\lambda_{\min}(\bSigma_g)||\bx^*||^2 = 
\lambda_{\min}(\bSigma_g)||\bx^{(o)}||^2$} 
where the \mbox{$d$-variate} vector $\bx^*$ 
has the cells of $\bx^{(o)}$ in the positions 
of $o$ and zeroes in the other positions. 
Therefore $\lambda_{\min}(\bSigma_g^{(o)})
\geqslant \lambda_{\min}(\bSigma_g)$, so 
all $|\bSigma_g^{(o_i)}| > 0$ and all 
$\MD^2(\bx_{g,i},\bw_{g,i},\bmu_g,\bSigma_g)$ 
in~\eqref{eq:cellMCD} exist. The second 
constraint enforces that at least $h_g$ cells 
remain unflagged in each column, where by 
default $h_g = \lceil 0.75 n_g \rceil$.
The cellMCD method is available in the
\textsf{R} package \texttt{cellWise} 
\citep{cellWise}, and has been applied to 
cluster analysis by 
\cite{zaccaria2025,zaccaria2026}.

In the LDA setting, we estimate the shared 
scatter matrix $\bSigma$ by pooling. We
first compute the location estimate $\bhmu_g$
of each class $g$ as the coordinatewise 
mean of its data imputed by the Detect 
Deviating Cells method \citep{DDC}. 
Next, we compute $\bhSigma$ by applying
the cellMCD with fixed center $\bzero$
to the pooled set of all 
$\bx_{g,i} - \bhmu_{g}$\,, and we use  
$\bhSigma$ in each class. 
(Note that there are other ways to regularize
the $\bhSigma_g$ when some classes have too 
few members. For instance, we could apply
cellMCD to the classes that do have enough
members, and in the small classes use a 
weighted average of those $\bhSigma_g$ 
with weights $n_g$\,.)

%================================================
\subsection{Flagging cells in test data} 
\label{sec:CF}

Even with robust estimates of 
$(\bmu_g, \bSigma_g)$, 
the discriminant rules in~\eqref{eq:cqda} 
and~\eqref{eq:clda} cannot robustly classify new 
data points with cellwise contamination. That is 
because the discriminant rule~\eqref{eq:cqda} is
quadratic in $\bx$ and~\eqref{eq:clda} is linear
in $\bx$, so outlying cells in $\bx$ can have a
large effect. Therefore we need to detect outliers 
in new data when classifying them. For this 
purpose we propose a likelihood-based
\textit{cellflagger}.

Given $(\bhmu_g, \bhSigma_g)$ and a single 
point $\bx$, cellflagger obtains a vector 
$\bw_g$ by minimizing the penalized objective
\begin{equation} \label{eq:CFobj}
  \mathcal{Q}\big(\bw \big) 
  = \log\big|\bhSigma_g^{(o(\bw))}\big| 
  + |o(\bw)| \log(2\pi) 
  + \MD^2 \big(\bx,\bw,\bhmu_g,\bhSigma_g\big) 
  + \sum_{j\mbox{\scriptsize{ in }} m} q_{gj}
\end{equation}
that is a part of the cellMCD 
objective~\eqref{eq:cellMCD} but now applied
to $\bx$. Flagging a previously unflagged 
cell $x_j$ decreases the objective by
\begin{align} \label{eq:deltaj}
\Delta_{gj} &= \mathcal{Q}(\bw : w_j = 1)
  - \mathcal{Q}(\bw : w_j = 0) \notag \\
& = \log\big|\bhSigma_g^{(o)}\big| 
  - \log\big|\bhSigma_g^{(o-j)}\big| 
  + \log(2\pi) + \MD^2(x_j, 
  \widehat{x}_{gj},C_{gj}) - q_{gj} \notag \\
& = \log C_{gj} + \log (2\pi) + \frac{(x_j - 
  \widehat{x}_{gj})^2}{C_{gj}} - q_{gj}\;, 
\end{align}
where $o-j$ denotes the set 
$o \setminus \{j\}$. Here 
$\widehat{x}_{gj} = \hmu_{gj} + 
 (\bhSigma_g)_{j,o-j} 
 (\bhSigma_g^{(o-j)})^{-1} 
 (\bx_{o-j} - (\bhmu_g)_{o-j})$
is the conditional expectation of cell $x_j$ 
and $C_{gj} = (\bhSigma_g)_{j,j} 
 - (\bhSigma_g)_{j,o-j} (\bhSigma_g^{(o-j)})^{-1} 
  (\bhSigma_g)_{o-j, j}$
is the conditional variance of $x_j$\,. 
The squared Mahalanobis distance of the 
univariate $x_j$ from the center 
$\widehat x_{gj}$ relative to the
variance $C_{gj}$ is simply 
$(x_j - \widehat x_{gj})^2/C_{gj}$\,. 
Cellflagger iteratively flags cell 
$j^\star=\argmax_j \Delta_{gj}$ if 
$\Delta_{gj^\star}\geqslant 0$, updates $o$ 
accordingly, and repeats 
until $\max_j \Delta_{gj}<0$.

We need to set $q_{gj}$ so that $\Delta_{gj}$ 
effectively filters outlying cells. If a cell
is outlying, its squared standardized residual 
$(x_j - \widehat x_{gj})^2/C_{gj}$ 
in~\eqref{eq:deltaj} tends to be large. As a 
cutoff we can use the chi-squared quantile 
$\chi^2_{1,0.99}$ with one degree of freedom 
and probability $0.99$. We wish to flag a 
cell when its residual is too large, so we set
\begin{equation} \label{eq:qj}
  q_{gj} = \chi^2_{1,0.99} + \log(2\pi) + 
  \log ( C_{gj} )
\end{equation}
be default. From~\eqref{eq:deltaj} 
and~\eqref{eq:qj} we see
that a cell $x_j$ is flagged with respect 
to group $g$ iff it lies 
outside a robust tolerance interval around its 
conditional expectation $\widehat x_{gj}$ with 
coverage 99\%. 

%================================================
\subsection{Robust discriminant function}
\label{sec:discriminant_function}

Classifying a new case $\bx$ requires choosing
between $G$ classes. This creates a conundrum: 
for every class, the cellflagger will flag 
some cells of $\bx$ until the remainder of 
$\bx$ is not outlying for that class, hence 
several classes may appear to fit $\bx$. To 
resolve this we will define a robust 
discriminant function based on the likelihood 
of $\bx$ in a contamination model. For any
potential class, it will naturally penalize 
based on how many cells of $\bx$ need to be 
flagged before it looks likely to belong to 
that class, and on how outlying the flagged 
cells are from that class.

For this novel situation we need to create a 
suitable cellwise contamination model. The new
model generates potentially contaminated data 
points as i.i.d.\ realizations of a $d$-variate
random vector $\cX$ of the form
\begin{align}\label{eq:cellwisecont}
 \cX &= \cW \odot \cY 
 + (\bone_d -\cW) \odot \cZ 
 \quad \mbox{where} \nonumber \\
 \cY \cond (\mathcal{G} = g) &\sim \N(\bmu_g, 
\bSigma_g) \quad \mbox{ for } 
 g=1,\ldots,G \nonumber \\
  (\cZ)_j \cond (\mathcal{G} = g) &\sim 
  \Lap(\mu_{gj}, \alpha_{gj})
  \quad \mbox{ for } j=1,\ldots,d \\
  (\cW)_j \cond (\mathcal{G} = g) &\sim 
  \Bern(1 - p_{gj}) \quad \mbox{ for }
  j=1,\ldots,d \nonumber \\
  (\cW, \cY, \cZ)\;
 &\mbox{are mutually independent conditional on }
 \mathcal{G}\,. \nonumber
\end{align}
Here $\odot$ denotes the Hadamard product that
multiplies vectors entry by entry. 
The \mbox{$d$-variate} random vector 
$\cY$ is the regular term, 
$\cZ$ is the contamination term with independent 
coordinates, and $\cW$ is binary with 
independent coordinates. The distribution
$\Lap(\mu_{gj}, \alpha_{gj})$ has density 
\mbox{$\exp(-|z_j - \mu_{gj}|/\alpha_{gj})/
(2\alpha_{gj})$}.

This model says that a cell at feature $j$ is 
outlying for class $g$ with probability $p_{gj}$\,. 
The values in outlying cells are independent for 
each feature-class pair, and they follow a 
heavier tailed distribution than the clean cells. 
The model~\eqref{eq:cellwisecont} extends that 
of \citet{alqallaf2009} to several classes, and 
adds a distributional model on $\cZ$, that we 
need to perform out-of-sample predictions on 
contaminated data.

Under the contamination model 
\eqref{eq:cellwisecont}, the joint likelihood 
of $\cX$ and $\cW$ for a given class $g$ 
can be factorized as
\begin{equation} \label{eq:XW|G}
f_{(\cX, \cW) \cond \cG}((\bx, \bw) \cond g) 
 = f_{\cW \cond \cG}(\bw \cond g) 
   f_{\cX \cond (\cW, \cG)}(\bx \cond (\bw, g)).
\end{equation}
Writing $o(\bw) = \{ j: w_j = 1 \}$ and 
$m(\bw) = \{j: w_j = 0\}$, the contamination model 
implies $\cX^{(o(\bw))} = \cY^{(o(\bw))}$ and 
$\cX^{(m(\bw))} = \cZ^{(m(\bw))}$, hence the
second factor becomes
\begin{equation} \label{eq:X|WG}
f_{\cX \cond (\cW, \cG)}(\bx \cond (\bw, g)) = 
f_{\cY^{(o(\bw))}\cond \cG}(\bx^{(o(\bw))}\cond g) \;
f_{\cZ^{(m(\bw))}\cond \cG}(\bx^{(m(\bw))}\cond g).
\end{equation}
Combining these equations yields
\begin{align} \label{eq:density}
f_{(\cX, \cW) \cond \cG} ((\bx, \bw) \cond g)
&= f_{\cW \cond \cG}(\bw \cond g) \;
  f_{\cY^{(o(\bw))}\cond \cG}(\bx^{(o(\bw))}
    \cond g) \;
  f_{\cZ^{(m(\bw))}\cond \cG}(\bx^{(m(\bw))}
  \cond g) \nonumber \\
&= \prod_{j=1}^d \big[ (1 - p_{gj})^{w_j}
   p_{gj}^{(1- w_j)} \big] \;\phi_{(o(\bw))} 
   \big(\bx^{(o(\bw))}\cond (\bmu_g^{(o(\bw))}, 
   \bSigma_g^{(o(\bw))})\big) \nonumber \\
&\quad \prod_{j \mbox{\scriptsize{ in }}m(\bw)} 
 \frac{\exp(-|x_j-\mu_{gj}|/\alpha_{gj})}
 {2\alpha_{gj}}\,,
\end{align}
where $\phi_{(o(\bw))}$ is the multivariate 
normal density over the features in $o(\bw)$.
Multiplying this density by the estimated prior 
probability $\hpi_g = n_g/n$ of class $g$, and 
taking the logarithm, yields the robust 
discriminant function
\begin{align} \label{eq:deltag}
 \delta_g&(\bx, \bw_g) := \log \hpi_g +
 \log \phi_{( o(\bw_g))} (\bx^{( o(\bw_g))} 
 \cond (\bmu_g^{( o(\bw_g))}, 
 \bSigma_g^{( o(\bw_g))}))
 \,+ \\
 &\sum_{j=1}^d \Big[ w_{gj} \log(1 - p_{gj}) 
 + (1- w_{gj})\log p_{gj} \Big]
 + \sum_{j=1}^d (1-w_{gj}) \log \Big( 
 \frac{\exp (-|x_j - \mu_{gj}|/ 
 \alpha_{gj})}{2 \alpha_{gj}}\Big).\nonumber 
\end{align}
When computing the discriminant function
for a new case $\bx$ and a class $g$ we first
determine $\bw_g$ by the cellflagger for
class $g$. Then we assign $\bx$ to the 
class with highest $\delta_g(\bx,\bw_g)$.

\begin{figure}[H] 
\centering
\includegraphics[width=0.55\linewidth]
  {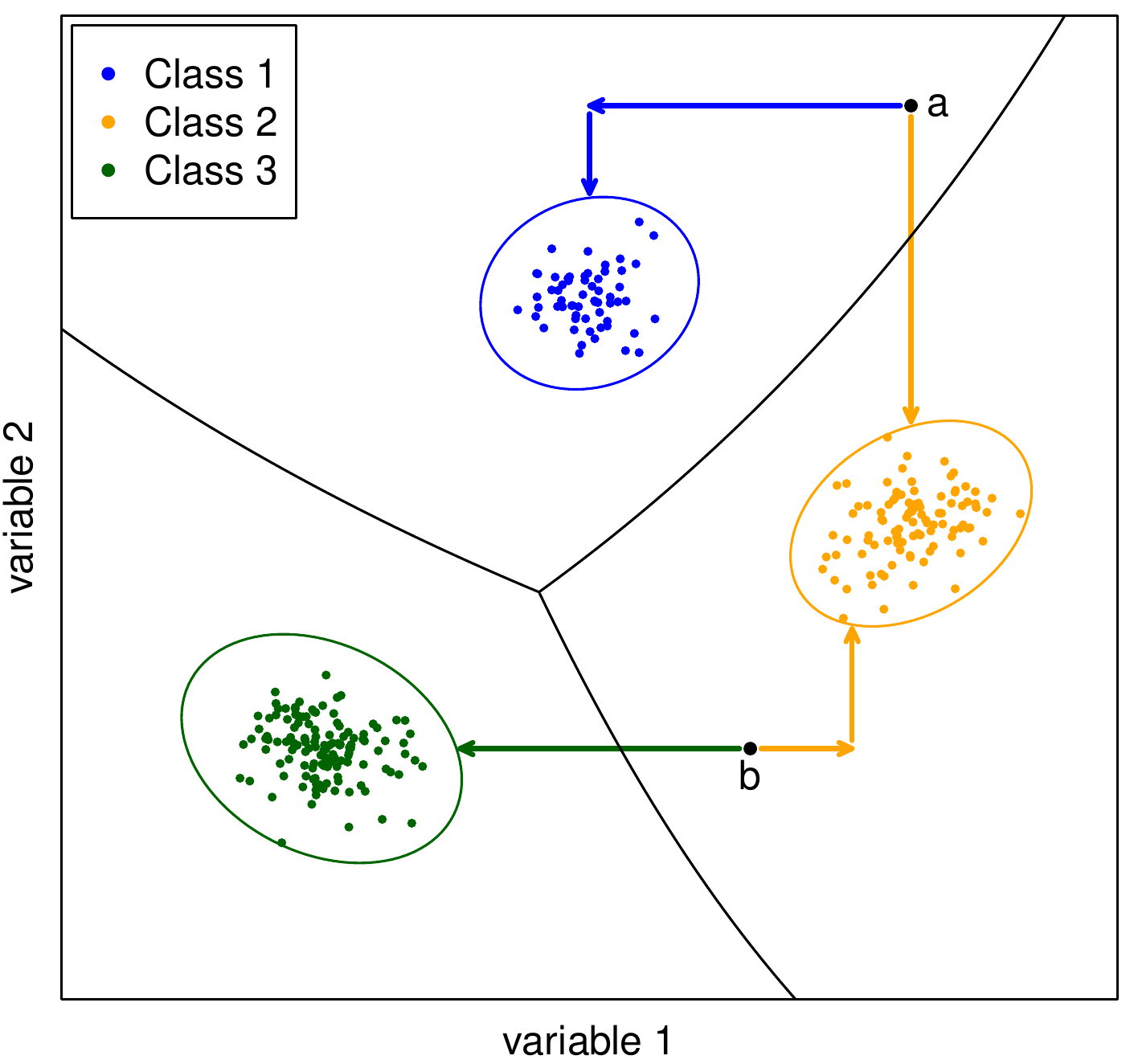}
\caption{Assigning
point \texttt{a} to class 1 yields two big
negative contributions (blue arrows) in
the discriminant rule~\eqref{eq:deltag},
whereas assigning it to class 2 yields only
one such term (orange arrow). Assigning
point \textbf{b} to class 2 yields two
such terms, against only one for class 3 
(green arrow).}
\label{fig:toy1_b}
\end{figure}

The robust discriminant 
rule~\eqref{eq:deltag} is the usual one, 
plus two new terms. We illustrate their 
roles with two examples. The first is the
toy example of Figure~\ref{fig:toy1_a}.
The classical rule would assign case 
\textbf{a} to class 1. However, suppose
that we know that class 1 has only few
outlying cells, that is, $p_{11}$ and
$p_{12}$ are tiny, whereas class 2 has
a higher probability $p_{22}$ of an
outlying cell in its second variable.
Then the new term
$\sum_{j=1}^2 [ w_{gj} \log(1 - p_{gj}) 
 + (1- w_{gj})\log p_{gj}]$
is very low for class $g=1$, since
$w_{11} = 0 = w_{12}$\,, and 
$\log p_{g1}$ and $\log p_{g2}$ are big 
negative values. For $g=2$ we find 
instead $w_{21} = 1, w_{22} = 0$ and
$\log p_{g2}$ is less extreme, so
point \textbf{a} can be assigned to 
class 2. In Figure~\ref{fig:toy1_b} we 
see that we would only need to move one 
cell of \textbf{a} to place it in class 2,
whereas the blue arrows indicate that we 
would need to move both cells of 
\textbf{a} to get to class 1.
Analogously, \textbf{b} can go to class 3
with only $w_{31}$ being zero (green
arrow), whereas placing it in class 2 
would require both $w_{21}$ and $w_{22}$ 
to be zero.

\begin{figure}[!ht]
\centering
\includegraphics[width=0.8\linewidth]
  {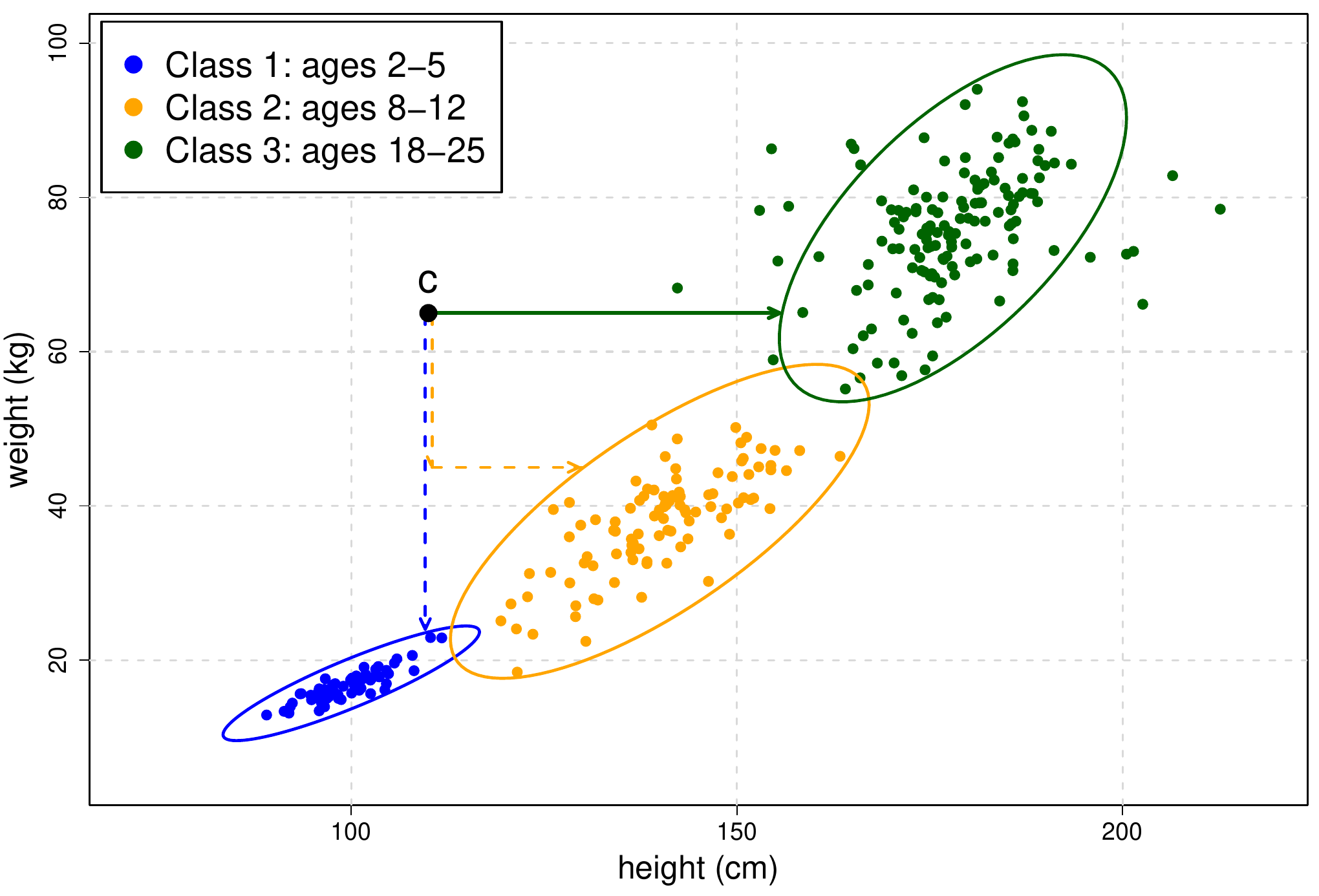}
\caption{Illustration of how the new
discriminant function~\eqref{eq:deltag}
assigns the out-of-sample point \textbf{c}
to class $g=3$, that has a higher spread 
$\alpha_{gj}$ of cellwise outliers in 
the variable Height ($j=1$).}
\label{fig:toy2}
\end{figure}

For the role of the second new term 
in~\eqref{eq:deltag}
look at Figure~\ref{fig:toy2}. The 
variables are height and weight of males
belonging to three age groups: 2--5 years
old in class 1, 8--12 years in class 2, 
and 18--25 years in class 3. The 
reported height of person \textbf{c} is
not particularly unusual, and neither is
his weight, but the combination is quite
unlikely. This may be due to an outlying
cell, but is it in height or in weight?
Now suppose that the Laplace parameter
$\alpha_{31}$ is large, whereas the other
five $\alpha_{gj}$ are tiny. That is,
the spread of cellwise outliers in the
height of the class of adults is large,
whereas all the other spreads are small.
For class 3 the final term
$\sum_{j=1}^d (1-w_{gj}) 
 \log ( (\exp (-|x_j - \mu_{gj}|/ 
 \alpha_{gj}))/(2 \alpha_{gj}))$ of 
the discriminant rule~\eqref{eq:deltag} has 
$w_{31} = 0, w_{32}=1$, and it is rather 
small due to the high $\alpha_{31}$\,. 
For classes 1 and 2 the term takes on
a large negative value due to their tiny 
$\alpha_{gj}$\,, so overall the value
of the discriminant rule is highest
for class 3. The method thus assigns 
point \textbf{c} to class 3.

An important benefit of deriving the 
discriminant rule~\eqref{eq:deltag} from the 
cellwise contamination 
model~\eqref{eq:cellwisecont} is its natural 
handling of missing data. If a case 
$\bx$ contains missing entries (NA's), we fix
$w_j=0$ in these cells, prior to the application 
of the cellflagger. We then permanently exclude 
those $j$ from the index set $o(\bw)$ of clean 
cells, and from both sums over $j$ in the 
discriminant function~\eqref{eq:deltag}.

We now look at the situation where the new case
$\bx$ does not belong to any of the classes, that
is, it is a casewise outlier for all classes. 
The above method will predict some class $g$, 
with the class-specific cellflagger output 
$\bw_g$\,. We will flag $\bx$ as a casewise 
outlier if the number of flagged cells 
$|m(\bw_g)|$ is larger than or equal to 
$d/2$, or when the Mahalanobis distance
$\MD^2(\bx,\bw_g,\bhmu_g,\bhSigma_g)$ of the
remaining cells exceeds the $0.99$ quantile of 
the $\chi^2$ distribution with 
$\sum_{j=1}^d w_j$ 
degrees of freedom. Such overall casewise
outliers are then assigned to a new class 
with the number 0.

%========================================
\subsection{Parameter estimation}

Our modeling choices for $\cW$ and $\cZ$ are 
motivated by simplicity and transparency. 
Contamination
at a cell is an on/off event, making 
Bernoulli indicators a natural choice. Also, 
the resulting log-likelihood expression remains 
separable, and yields an explicit penalty on 
the number of flagged cells, that is analogous 
to the columnwise penalty terms in the 
objective~\eqref{eq:cellMCD} of cellMCD and 
the objective~\eqref{eq:CFobj} of the 
cellflagger. The Laplace model has heavier 
tails than the normal model, and for flagged 
cells it subtracts
$\sum_{j=1}^d (1-w_{gj}) 
 [|(x_j - \mu_{gj})/\alpha_{gj}|
 -\log(2\alpha_{gj})]$ from the 
log-likelihood~\eqref{eq:deltag}. The sum 
of the $|(x_j - \mu_{gj})/\alpha_{gj}|
|1-w_{gj}|$ is a weighted $L_1$ norm of
$\bone - \bw$, and the sum of the
$\log(2\alpha_{gj})|1-w_{gj}|$ is a 
weighted $L_0$ pseudonorm of $\bone - \bw$. 
Both are loosely connected to the
sparsity of $\bone - \bw$ that we 
saw in cellMCD and the cellflagger, where
the number of flagged cells was penalized.

We estimate the contamination parameters 
$(p_{gj}, \alpha_{gj})$ from flagged cells 
in the training data. That is, after computing 
the cellMCD estimates $(\bhmu_g, \bhSigma_g)$ 
of the class parameters, we apply the
cellflagger to each training case $\bx_{g,i}$ 
with its given class $g$ to obtain flags 
$\bw_{g,i}$\,. This step is necessary 
because the flagging patterns from cellMCD, 
obtained by joint optimization over $\bmu_g$, 
$\bSigma_g$ and $\bW_g$, may differ slightly from 
those of the cellflagger that takes $\bhmu_g$ 
and $\bhSigma_g$ 
as given. Moreover, this step ensures the same 
flagging in-sample and out-of-sample. In other 
words, if we encounter an in-sample $\bx_{g,i}$ in 
the test dataset, it will get the same 
$\bw=\bw_{g,i}$ that it had in the training data.
This makes the approach internally consistent.

We first estimate the Bernoulli parameters
$p_{gj}$\,. For class $g$ and feature $j$, we
denote the number of flagged cells by
$m_{gj} = \sum_{i=1}^{n_g} (1 - w_{g,ij})$. We 
estimate the contamination probability from
the empirical frequency of the flagged cells:
\begin{equation} \label{eq:phat}
  \widehat{p}_{gj} = \max\left(0.01,
  \frac{m_{gj}}{n_g} \right),
\end{equation}
where the floor $0.01$ matches the tolerance 
level of the cellflagger. Even when no outlying 
cells are generated $(p_{gj} = 0)$, the 
cellflagger will 
still flag approximately 1\% of cells due to 
its use of the quantile $\chi^2_{1,0.99}$ by
default. Therefore, we set the minimal value of 
$\widehat p_{gj}$ to 0.01 in order to account 
for the baseline flagging rate. This was found 
to work well in simulation.

Next, we estimate the Laplace parameter
$\alpha_{gj}$\,. Denote the sum of absolute 
deviations over flagged cells as
\[
S_{gj} = \sum_{i=1}^{n_g} (1-w_{g,ij})
         |x_{g,ij} - \hmu_{gj}|\,.
\]
The maximum likelihood estimator
$\halpha^{\text{MLE}}_{gj} = \frac{S_{gj}}{m_{gj}}$
provides reliable estimates when sufficiently 
many cells are flagged. When no cells are flagged, 
we set a default scale estimate by matching the 0.995 
quantiles of the Laplace and the normal distributions. 
This yields the default scale estimate
\[
 \halpha^{(0)}_{gj} := \frac{z_{0.995}}{\log(100)}
 \sqrt{\frac{1}{(\widehat\bSigma_g^{-1})_{jj}}} 
 \approx 0.56 \sqrt{\frac{1}
 {(\widehat\bSigma_g^{-1})_{jj}}}\;.
\]
To stabilize the estimate of $\alpha_{gj}$ when 
$m_{gj}$ is nonzero but small, our final estimate
is a weighted average of the default estimate and 
the MLE:
\begin{equation} \label{eq:alphahat}
\widehat{\alpha}_{gj} = 
  \frac{\tau \alpha^{(0)}_{gj} +
  m_{gj}\halpha^{\text{MLE}}_{gj}}{\tau + m_{gj}} 
  = \frac{\tau}{\tau + m_{gj}}\halpha^{(0)}_{gj} 
  + \frac{m_{gj}}{\tau + m_{gj}}
  \halpha^{\text{MLE}}_{gj}\,.
\end{equation}
This equals $\halpha^{(0)}_{gj}$ when 
$m_{gj} = 0$ and tends to 
$\halpha^{\text{MLE}}_{gj}$ for increasing
$m_{gj}$\,. We set 
\mbox{$\tau = \min(1, \frac{n_g}{100})$} to 
match the false positive rate of the 
cellflagger, that flags 
approximately 1\% of clean cells. Capping 
$\tau$ at 1 serves the following purpose.
We know that when 10 or more cells are flagged 
the MLE is stable, and for 
$m_{gj} \geqslant 10$ the weight of the default 
scale $\halpha^{(0)}_{gj}$ indeed gets small 
enough due to $\tau/(\tau + m_{gj}) \leqslant 
 1/(1+m_{gj}) \leqslant 1/11 < 0.1$\,. 

In the quadratic discriminant analysis setting 
we will use the name cellQDA for the entire 
procedure that includes estimation of all
parameters on the training data, as well as the
robust discriminant function for out-of-sample
data. Its linear counterpart will be denoted 
cellLDA.

%============================================
\section{Simulations} \label{sec:sim}

We will evaluate the performance of discriminant 
analysis methods on simulated data with cellwise 
and/or casewise 
contamination. Clean data are generated from a 
normal mixture of $G=3$ groups. The training 
and test cases in group $g$ are sampled from 
$\N (\bmu_g, \bSigma_g)$. For each class, we 
generate 200 training cases and 200 test cases.
The simulations are carried out in dimensions
$d=5$ and $d=20$. 

%============================================
\subsection{Quadratic Discriminant Analysis}
\label{sec:simulQDA}

For the true $\bSigma_g$ we use correlation
matrices with decaying off-diagonal entries, 
such as the A09 correlation matrix that has the 
entries $(\bSigma_g)_{ij} = (-0.9)^{| i - j |}$.
By replacing $0.9$ by other multiples of $0.1$ 
we obtain analogous matrices like A08 and A04. 
We consider two \mbox{scenarios}, with high and 
low correlation. In the high-correlation scenario 
we set \mbox{$(\bSigma_1, \bSigma_2, \bSigma_3) 
= \text{(A09, A08, A07)}$}, whereas in the 
low-correlation scenario we use\linebreak 
\mbox{$(\bSigma_1, \bSigma_2, \bSigma_3) =
\text{(A06, A04, A02)}$}. We choose the true 
class centers  as follows: $\bmu_1 = \bzero$, 
\mbox{$\bmu_2 = \bone_d$\,,} and 
$\bmu_3 = \bv$, where $\bone_d$ is the 
$d \times 1$ column vector of ones and 
$\bv = (2, -2, 2, -2, \dots)^\top$ 
is a vector with alternating signs. 

We consider three contamination types.
In the cellwise outlier scenario we replace 
$10\%$ of random cells $x_{ij}$ by 
$\mu_{gj}+\gamma\sqrt{(\bSigma_g)_{jj}}$\,, 
where $\gamma$ varies from 0 to 10. When 
$\gamma = 0$, we do not contaminate the data.
In the casewise outlier scenario, $10\%$ of 
the cases in class $g$ are generated from 
$\N(\bmu_g + \frac{\gamma}{2}(\bmu_{g'}
- \bmu_g), \bSigma_g)$ where $g'$ denotes 
the next class in the cyclic ordering 
$1\rightarrow 2\rightarrow 3\rightarrow 1$.
In the mixed contamination scenario, we 
contaminate $5\%$ of the cells and $5\%$ of 
the cases. The contaminated data are 
generated by the function 
\texttt{generateData()} in the \textsf{R} 
package \texttt{cellWise}. 
Each combination of scenarios is replicated 
20 times.

As a baseline we include classical QDA, 
denoted as CQDA, which uses the mean and 
empirical covariance matrix of each class 
in its discriminant rule. To the best of 
our knowledge, \cite{aerts2017} is the 
only prior work on cellwise robust QDA.
They consider several approaches, and for
quadratic discriminant analysis they
find that the best performing one  
estimates the $\bSigma_g$ by the cellwise 
robust graphical lasso. We will refer to 
their QDA method as GLQDA, where GL stands
for Graphical Lasso. We also 
include the casewise robust method 
of \citet{RQDA}, denoted RQDA, that 
estimates the $\bmu_g$ and $\bSigma_g$ by 
the casewise MCD using the algorithm of 
\cite{FastMCD}. RQDA is able to flag 
casewise outliers in the test data, and 
excludes them from the classification by 
putting them in the separate class 0.

%=========================================
\subsubsection{Performance with clean 
  test data}
\label{sec:sim_no_casewise}

We evaluate all four methods under the three 
scenarios described above, where the training 
data contains cellwise outliers only, or 
casewise outliers only, or the combination 
of both. We then consider three different 
settings for the test data: without any 
outliers (clean), with cellwise outliers, 
and with casewise outliers. 

We start by comparing the four methods for
clean \textit{test} data. The comparison
is a bit tricky, since CQDA and GLQDA
assign the test cases to a class in
$\{1,2,3\}$, whereas RQDA and cellQDA will
by default assign them to a class in
$\{0,1,2,3\}$ where 0 is the extra
class for flagged outlying cases. Since 
the accuracy measure 
should be the same for all methods in the 
comparison, for now we turn off the 
casewise outlier detection mechanisms of 
RQDA and cellQDA. This forces them to 
assign all test cases to the same three 
classes $\{1,2,3\}$.

\begin{figure}[!ht]
\centering
\includegraphics[width=0.99\linewidth]
  {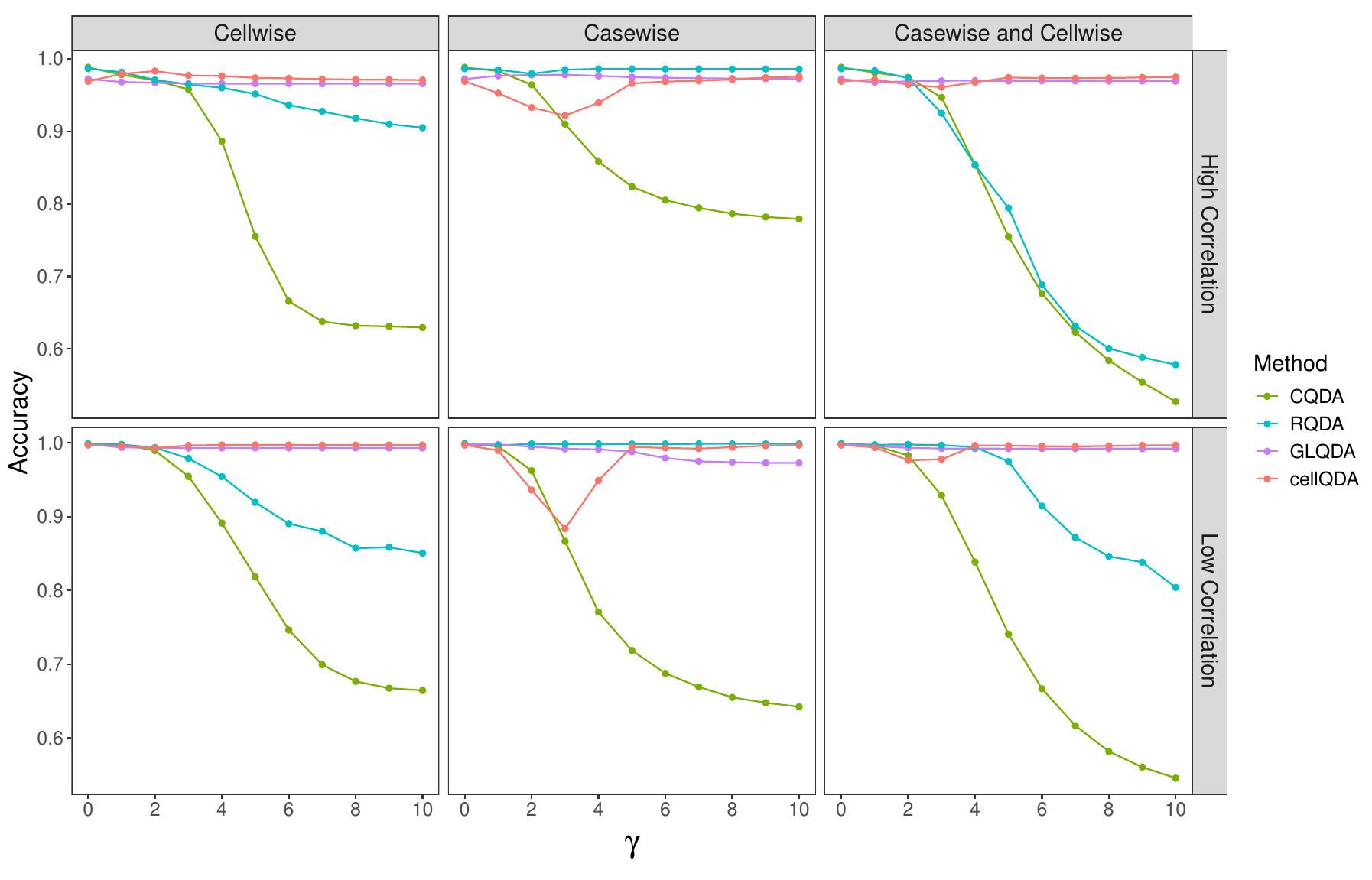}\\
\vspace{-3mm}
\caption{Accuracy of four QDA methods 
on 3 classes in 20 dimensions, with $n=200$. 
In the top row the classes have high within 
correlations, and the bottom row has low within 
correlations. The first column uses training
data with 10\% of cellwise outliers only, the 
second with 10\% of casewise outliers only, and
the third with 5\% of cellwise and 5\% of 
casewise outliers. The parameter $\gamma$ on the
horizontal axis says how far away the outliers 
are. The accuracy is measured on test data that
contains no contamination.}
\label{fig:sim_clean_d20}
\end{figure}

Figure~\ref{fig:sim_clean_d20} shows the 
resulting accuracies as a 
function of $\gamma$ in dimension $d=20$, with 
the high-correlation scenario in the top row 
and the low-correlation scenario in the 
bottom row. For clean \textit{training} data
(labeled as $\gamma = 0$), all methods attain 
an accuracy close to 100\%. 
But for $\gamma>1$ some accuracies degrade. 
The performance depends a lot on the type of 
contamination. In the first and third columns 
we see that whenever cellwise outliers are 
present in the training data, cellQDA and
GLQDA are the only methods that can withstand 
increasing $\gamma$ values. In the middle column 
only casewise outliers are present in the training
data, and then the casewise robust method 
RQDA did very well, followed closely by cellQDA 
and GLQDA, but not CQDA. 

%=============================================
\subsubsection{Performance with cellwise 
  outliers in the test set}
\label{sec:sim_cellwise}

With its clean test data, 
Figure~\ref{fig:sim_clean_d20} basically 
reflects the robustness and efficiency of 
the estimators of $\bmu_g$ and $\bSigma_g$ 
during training. 

\begin{figure}[!ht]
\centering
\includegraphics[width=0.99\linewidth]
{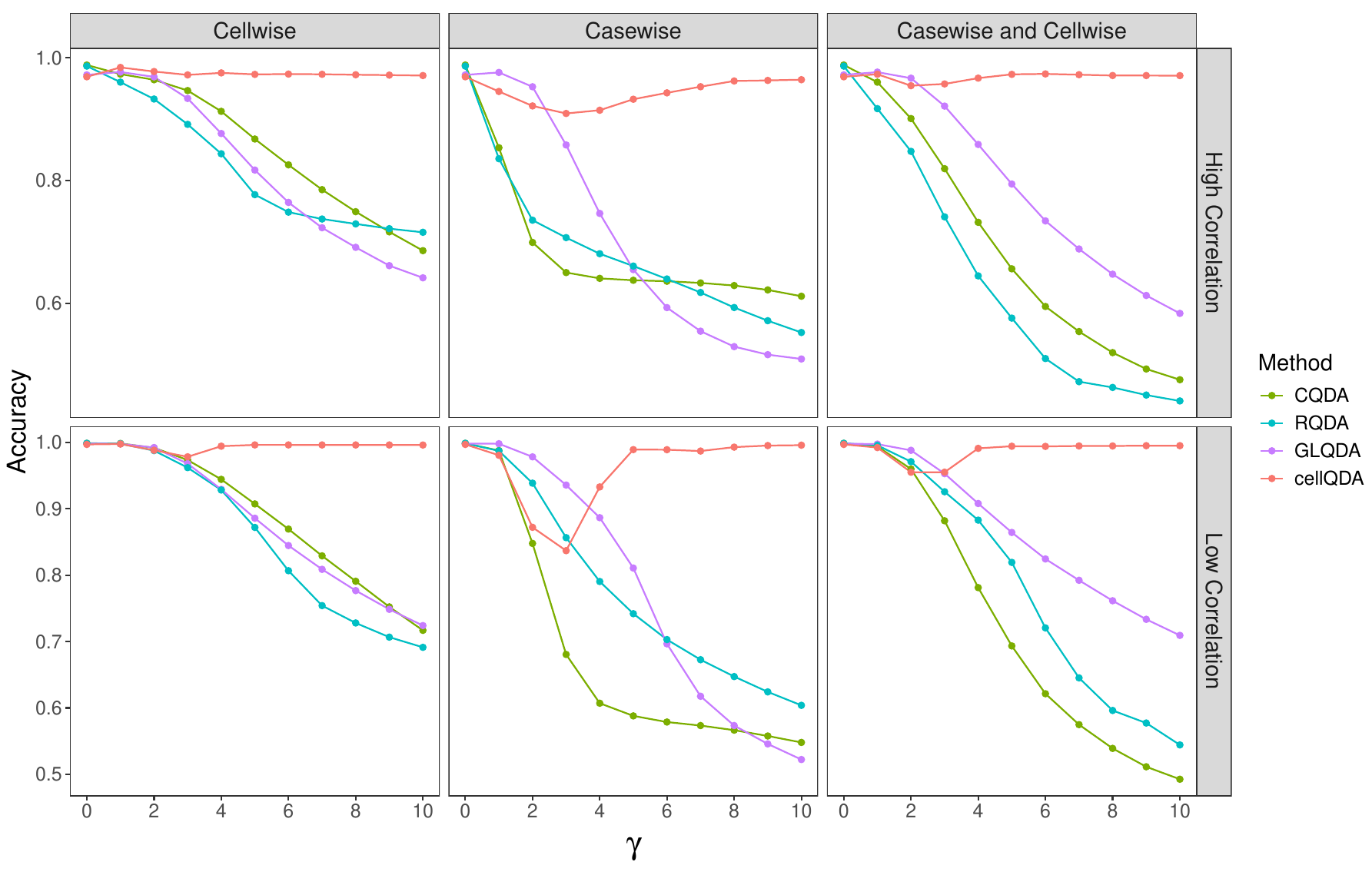}\\
\vspace{-3mm}
\caption{Comparison of the four QDA
methods on the same training data as in 
Figure~\ref{fig:sim_clean_d20} and displayed 
in the same way. The difference is in the
test data, that now contains 10\% of cellwise
outliers with the same $\gamma$ used in 
training.}
\label{fig:sim_cellwise_d20}
\end{figure}

However, it often happens that the test data 
are contaminated as well, so we need to study 
the performance in such situations also.
Figure~\ref{fig:sim_cellwise_d20} compares 
the methods when the test data has 10\% of
cellwise outliers only, with the same $\gamma$ 
values used in the training data. Note that
the training data as well as the estimates
of $\bmu_g$ and $\bSigma_g$ stay exactly
the same as before. We see that when cellwise 
contamination is introduced in the test set, 
the performance of all methods except cellQDA 
deteriorates substantially. When $\gamma$ 
increases, only cellQDA remains standing.
From Figure~\ref{fig:sim_clean_d20} we know
that RQDA and GLQDA did obtain accurate 
estimates of the class parameters $\bmu_g$ 
and $\bSigma_g$\,. Their decline here
clearly demonstrates the importance of a
discriminant rule that accounts for 
contaminated cells, as only cellQDA has
that property.

%==============================================
\subsubsection{Performance with casewise
  outliers in the test set}
\label{sec:sim_casewise}

It is also possible that the test data 
contains casewise outliers. We now generate
test data with 10\% of casewise outliers 
with the same $\gamma$ as in training, and
without cellwise outliers.

\begin{figure}[!ht]
\centering
\includegraphics[width=0.99\linewidth]
  {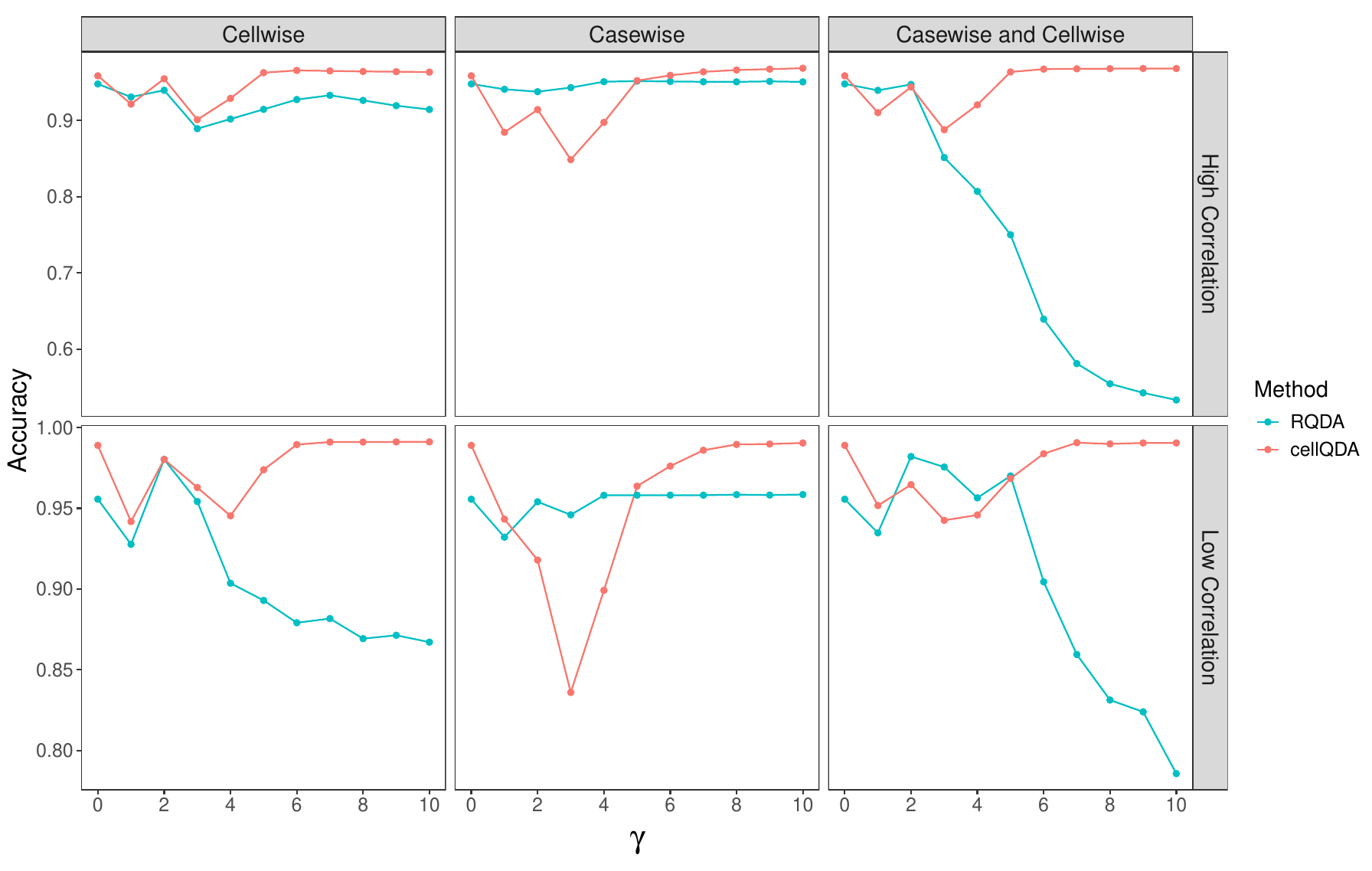}\\
\vspace{-3mm}
\caption{Comparison of RQDA and cellQDA on
the same training data as in 
Figure~\ref{fig:sim_clean_d20} and displayed 
in the same way. Now the test data contains 
a class 0 with 10\% of generated casewise 
outliers. The reported accuracy is of the 
assignments to the four classes 0, 1, 2, 
and 3.}
\label{fig:sim_casewise_d20}
\end{figure}

For the computation of the accuracy, we 
define the true class 0 as the set of 
generated casewise outliers. We switch 
the casewise detection rules of 
RQDA and cellQDA back on, so they can 
assign flagged cases to class 0. We  
compute the accuracy as the number of 
correct assignments to the classes in 
$\{0, 1, 2, 3\}$, divided by the size 
of the test set. Since CQDA and GLQDA 
don't assign cases to a class 0, they 
are absent in this comparison.

In the middle column we see that when both
the training data and the test data only
have casewise outliers, cellQDA does about
as well as RQDA that was designed for
casewise outliers. But when the training
data also contained cellwise outliers,
cellQDA outperformed RQDA.

%===========================================
\subsection{Linear Discriminant Analysis}
\label{sec:simulLDA}

Linear discriminant analysis assumes that all
classes were generated with the same scatter 
matrix $\bSigma$. In our simulation we use 
$\bSigma$ of type A09 in the high correlation
scenario, and of type A06 in the low 
correlation setting. The class centers
$\bmu_1$, $\bmu_2,$ and $\bmu_3$ are the same
as in QDA.

Our comparison includes 
classical LDA, denoted as CLDA. It first 
computes the mean $\bbar{\bmu}_g$ of each
class $g$, and then estimates the shared 
covariance matrix $\bSigma$ by the empirical 
covariance of the pooled centered data
$\{\bx_i-\bbar{\bmu}_{g_i}\,;\;i=1,\ldots,n\}$.
The LDA alternative of GLQDA first
computes robust estimates $\bhmu_g$ and then 
estimates $\bSigma$ by applying the cellwise 
robust graphical lasso to the
$\bx_i - \bhmu_{g_i}$\,.
We will refer to this method as GLLDA.
We also include a casewise robust linear 
discriminant analysis method of \citet{RQDA}, 
denoted as RLDA.
They describe three different approaches. 
We use the version that pools classes
in the same way as CLDA, cellLDA, and GLLDA.
First the estimates $\bhmu_g$ of the centers 
are obtained by the casewise MCD, and then 
the shared $\bSigma$ is estimated by applying
the casewise MCD to the $\bx_i - \bhmu_{g_i}$\,.
Afterward a reweighting step is carried out,
based on robust Mahalanobis distances, that
flags casewise outliers in the training data.
Next, the flagged cases are removed from the 
final estimates of the class centers, the 
shared covariance matrix, and the prior
probabilities of the classes. The resulting 
estimates are then plugged into the linear 
discriminant rule. As in RQDA, cases with 
sufficiently large robust Mahalanobis 
distance are assigned to class 0.

\begin{figure}[!ht]
\centering
\includegraphics[width=0.99\linewidth]
  {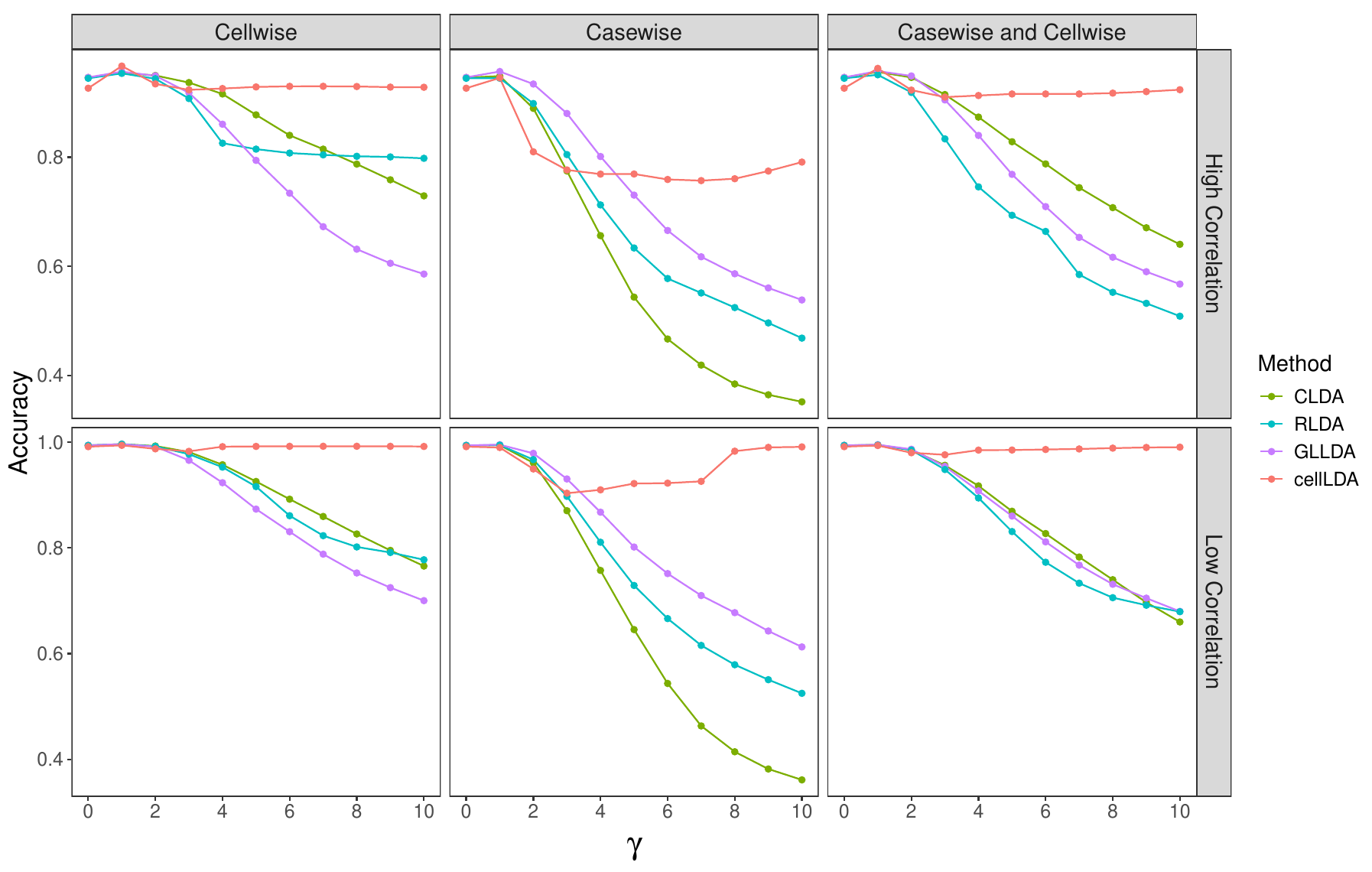}
\vspace{-3mm}
\caption{Comparison of four LDA methods on
three classes in 20 dimensions with $n=200$,
similar to Figure~\ref{fig:sim_cellwise_d20}.
Now the three classes were generated with the
same covariance matrix $\bSigma$, that has
high correlations in the top row and 
low correlations in the bottom row.}
\label{fig:LDA_d20}
\end{figure}

Figure~\ref{fig:LDA_d20} shows the accuracy 
when the test data is contaminated with 10\%
of cellwise outliers, in the same way as in
Figure~\ref{fig:sim_cellwise_d20}. Only cellLDA 
holds its own when $\gamma$ grows. CPCA  
already suffers during training, because the 
outliers affect the $\bm{\bar{\mu}}_g$ and
the empirical covariance matrix. The other
methods give more robust estimates of $\bmu$
and $\bSigma$, but during the out-of-sample
prediction only the cellLDA discriminant rule
avoids the effect of the outlying cells in
the test data.

All the simulations reported here are for
$d=20$ dimensions. The results for $d=5$ are
qualitatively similar and can be found in 
Section~\ref{supp:simul} of the Supplementary 
Material.

%============================================
\section{A real data example} 
\label{sec:realdata}

We will analyze a subset of the 
\texttt{nutrients\_branded} dataset that is 
publicly available in the R-package 
\texttt{robCompositions} \citep{templ2020}. 
It contains data on 804 different sweets sold 
in Switzerland. They are divided into four 
classes: ``Cookies and Biscuits'', 
``Milk-based ice cream'', 
``Cakes and tarts'', and ``Creams and 
puddings". It has 9 variables, describing 
nutritional contents: energy (kcal), proteins, 
water, carbohydrates, sugars, dietary fibers,
total fat, saturated fatty acids, and salt.
The dataset was analyzed earlier by 
\cite{classmap} using support vector machines.
That paper found some misclassified cases and 
casewise outliers, and afterward interpreted
them as being due to some atypical contents.
Here we will instead flag cellwise outliers 
that directly indicate unusual contents, such 
as unexpectedly high salt.

\begin{figure}[H]
\vspace{2mm}
\centering
\includegraphics[width=0.7\linewidth]
  {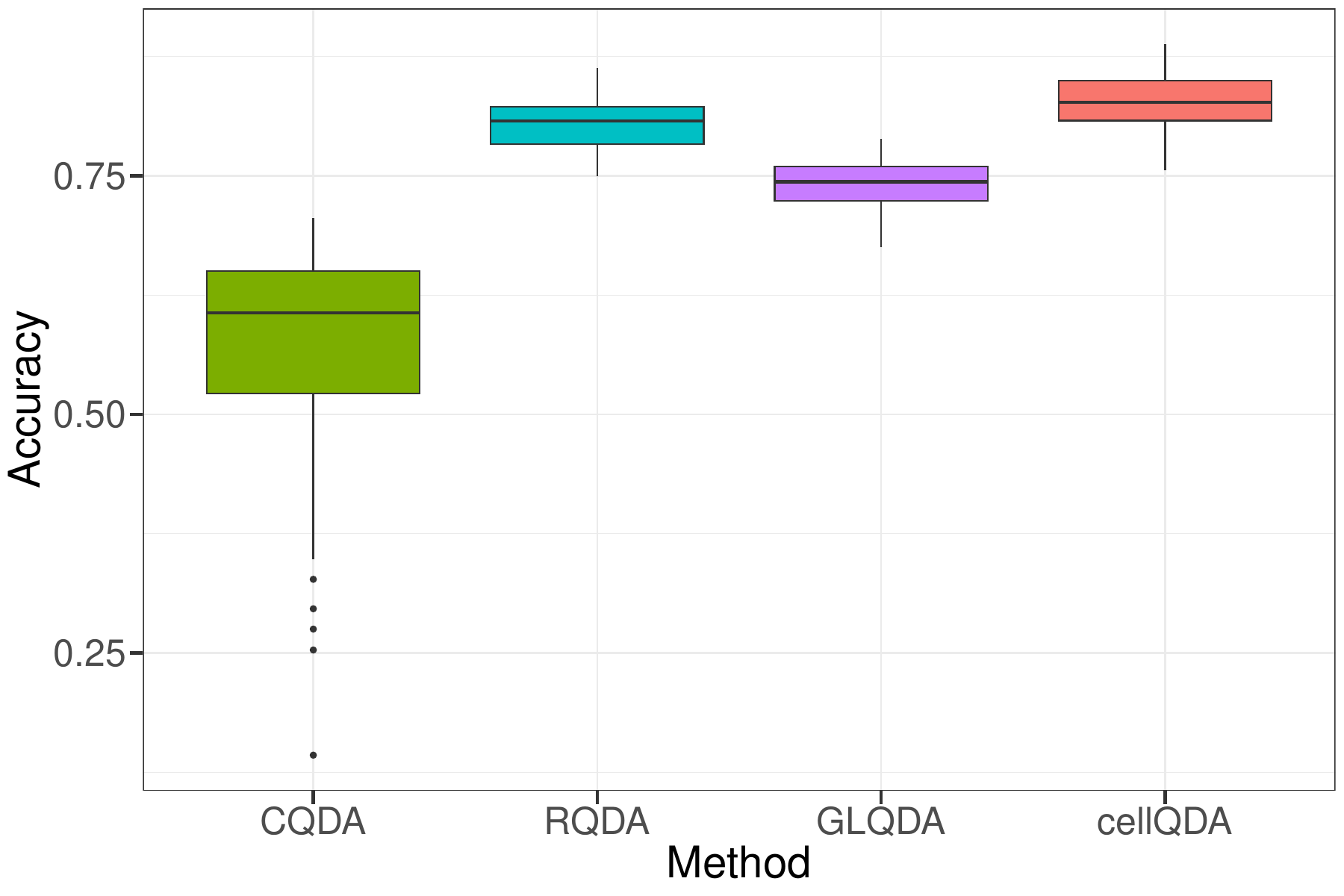}\\
\caption{Boxplot of cross-validated accuracies.}
\label{fig:sweets_boxplot}
\end{figure}

We computed the out-of-sample classification 
accuracy of CQDA, RQDA, GLQDA, and cellQDA by
5-fold cross-validation, replicated 10 times
on different random folds. The average accuracy 
of cellQDA was 0.832, and exceeds those of 
CQDA (0.569), RQDA (0.805), and GLQDA (0.767). 
Figure~\ref{fig:sweets_boxplot} shows boxplots 
of the accuracies on the 10 times 5 test sets.
Classical CQDA has the highest variability here.

\begin{figure}[!ht]
\centering
\includegraphics[width=1.0\linewidth]
  {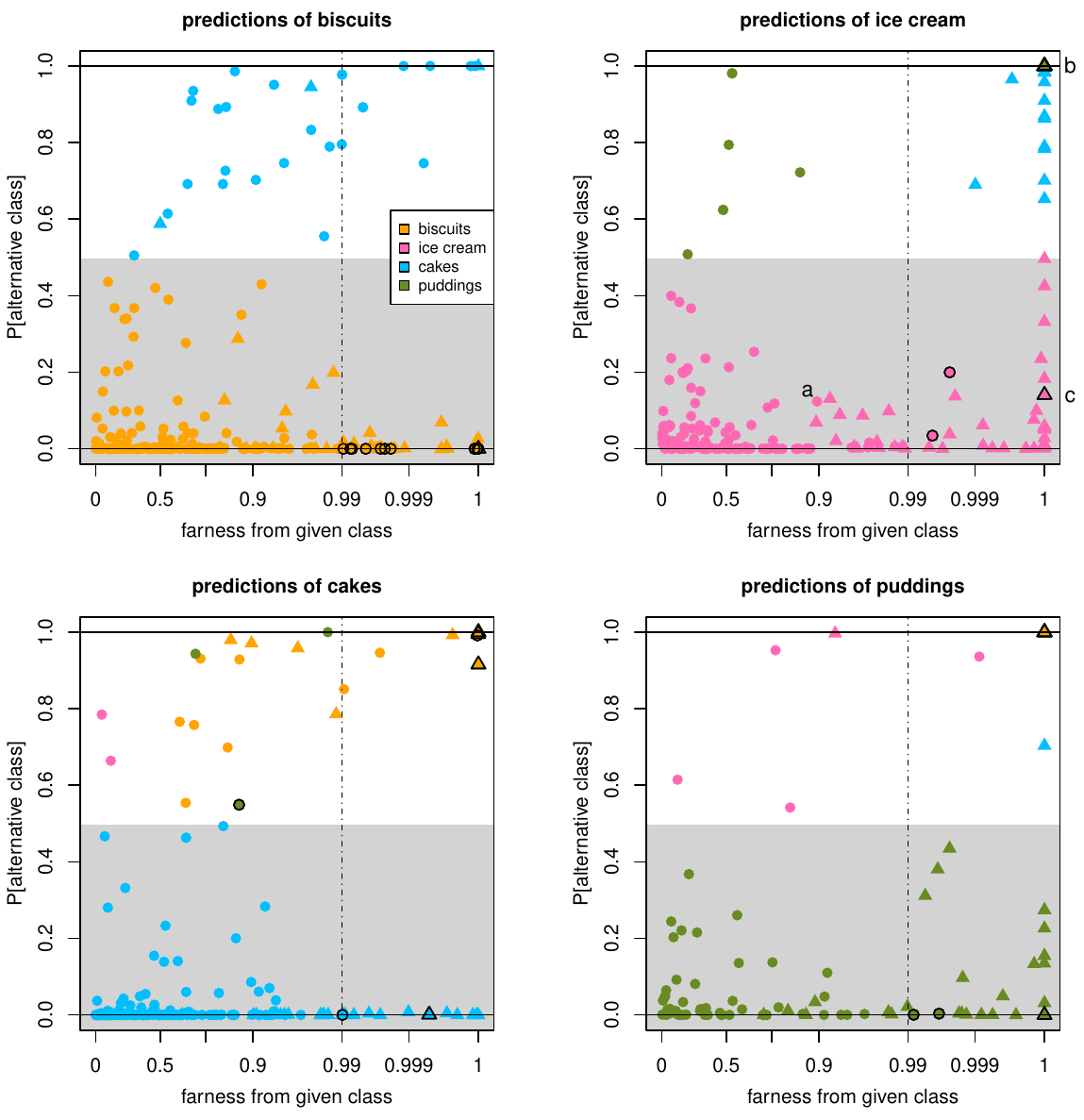}\\
\vspace{-2mm}
\caption{Class maps of the cellQDA 
classification of the sweets data, one
map per given class. The points are colored 
according to their predicted class.
Points with a black boundary were put in 
the class 0 of the flagged cases. 
Triangles indicate cases with at 
least one flagged cell.}
\label{fig:classmap_cellQDA}
\end{figure}

We can visualize the cellQDA classification
by classmaps, as introduced in \cite{classmap}.
Each of the 4 classes has its own classmap.
Let us focus on the classmap of the products
labeled as ice cream, in the top right panel
of Figure~\ref{fig:classmap_cellQDA}. For
each ice cream product, the horizontal axis 
shows its robust Mahalanobis distance to the 
center of the ice cream class. These distances 
are labeled as their cumulative distribution 
values for the $\sqrt{\chi^2_d}$ 
distribution with $d=9$ degrees of freedom.
There is a vertical cutoff line at its
99\% quantile. Moreover, the horizontal axis 
is not equispaced. Instead, the quantiles are
spaced like those of the standard univariate 
normal restricted to the interval $[0,4]$.

On the vertical axis of the classmap we
see another quantity ranging from 0 to 1.
This is the Probability of the Alternative
Class (PAC). It is defined as follows. The 
posterior probability of a case $i$
belonging to class $g$ follows 
from~\eqref{eq:deltag}:
\begin{align*} 
 \hp(i,g) &= \hpi_g\,
 \phi_{(o(\bw_i))} \Big(\bx_i^{(o(\bw_i))} 
 \cond (\bhmu_g^{( o(\bw_i))}, 
 \bhSigma_g^{(o(\bw_i))})\Big)\\
 &\quad \quad \prod_{j=1}^d \Big[
 (1 - \hp_{gj})^{w_{ij}}
 \hp_{gj}^{(1- w_{ij})} \Big] 
 \prod_{j=1}^d \Big( 
 \frac{\exp(-|x_{ij}-\hmu_{gj}|/
 \halpha_{gj})}{2\halpha_{gj}}
 \Big)^{(1-w_{ij})}\,.
\end{align*}
The given label of case $i$ is $g_i$\,, 
but that does not always mean that the 
posterior probability $\hp(i,g)$ is the 
highest at $g = g_i$. We define the
highest $\hp(i,g)$ attained at a class 
\textit{different from} $g_i$ as
\begin{equation*}
  \tp(i) := \max\{\hp(i,g)\,;\,
  g \neq g_i\}.
\end{equation*}
The class attaining this maximum can be seen
as the best alternative to class $g_i$\,.
When $\hp(i,g_i) > \tp(i)$ we know that $g_i$
attains the highest posterior probability in
case $i$, so the classifier will indeed 
assign $i$ to class $g_i$\,. However, when 
$\hp(i,g_i) < \tp(i)$ the MAP rule won't
assign object $i$ to class $g_i$ but to the
best alternative class, so case $i$ is 
misclassified. 

We now compute the PAC as the conditional
posterior probability of the best alternative
class, given that we can only choose between 
that class and $g_i$\,. This yields
\begin{equation}
  \PAC(i) := \frac{\tp(i)}
  {\hp(i,g_i) + \tp(i)} \,,
\end{equation}
and that is plotted on the vertical axis.
Note that when $\PAC(i) < 0.5$ the classifier
does predict the given class $g_i$\,,
whereas $\PAC(i) > 0.5$ means that it prefers
the best alternative class instead. This 
natural boundary at 0.5 is why the lower half 
of the classmap has a light grey background: 
that is the region where the classifier agrees 
with the given label. The upper half contains
the misclassified cases. All cases are colored
by their predicted class. So in the lower half
we only see pink points, the color of the ice
cream class in the legend box. Here the 
misclassified points are green (puddings) or 
blue (cakes). The points with a black 
boundary are those that were put in class 0 
after being assigned.

The original classmap only reflected casewise 
information. Here we add some cellwise 
information to it. Points without flagged 
cells are still plotted as round shapes, but 
now points with at least one flagged cell
get a triangular shape. We see that the 
triangles are often on the right hand side of 
the classmap, as far cellwise outliers can 
cause large Mahalanobis distances. Point (a) 
is round, and lies in the regular area with a 
small PAC and a distance below the cutoff. 
Point (b) is misclassified
as a pudding (green) with high 
conviction, as its PAC is close to 1. It is 
a casewise outlier with high distance, and 
it is a triangle so it has at least one 
outlying cell. Point (c) is also a casewise 
outlier with at least one outlying cell, but 
classified correctly as an ice cream. 

\begin{figure}[!ht]
\centering
\includegraphics[width=0.65\linewidth]
  {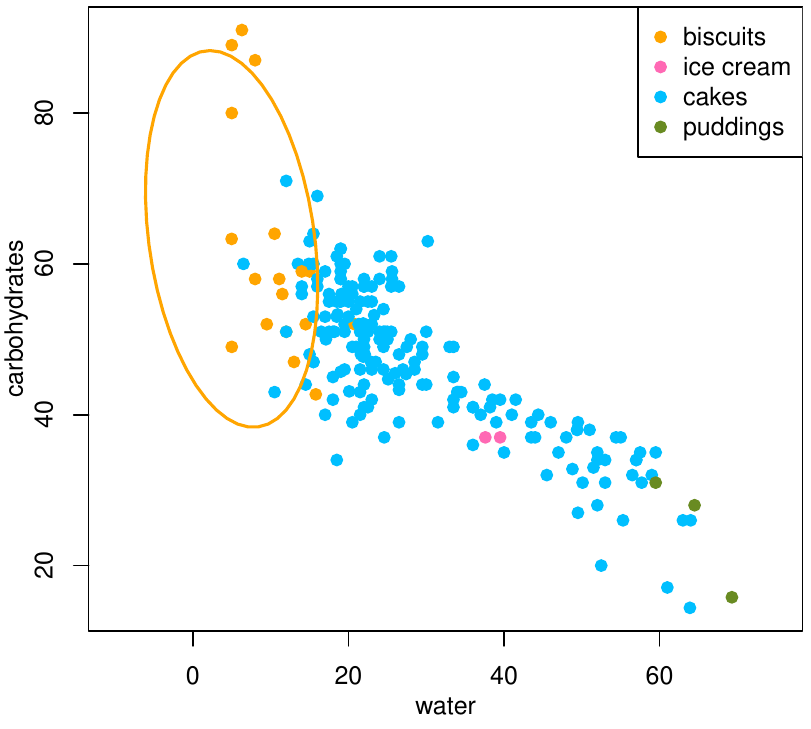}\\
\caption{Scatter plot of classification results 
of the cakes class. Datapoints are colored by 
their predicted class. The tolerance ellipse of
the biscuits class is shown in orange.}
\label{fig:cakes}
\end{figure}

In the bottom left panel of 
Figure~\ref{fig:classmap_cellQDA} we see that
many cakes are misclassified as orange points,
so biscuits. Figure~\ref{fig:cakes} shows 
the scatterplot of \texttt{carbohydrates}
versus \texttt{water} content. We see that
most of the cakes that are misclassified as
biscuits have very low water content, that is
more common for biscuits than for cakes. 
Most of these points correspond to cake mixes 
or fairly dry cakes based on nuts.

\begin{figure}[!ht]
\centering
\includegraphics[width=\linewidth]
  {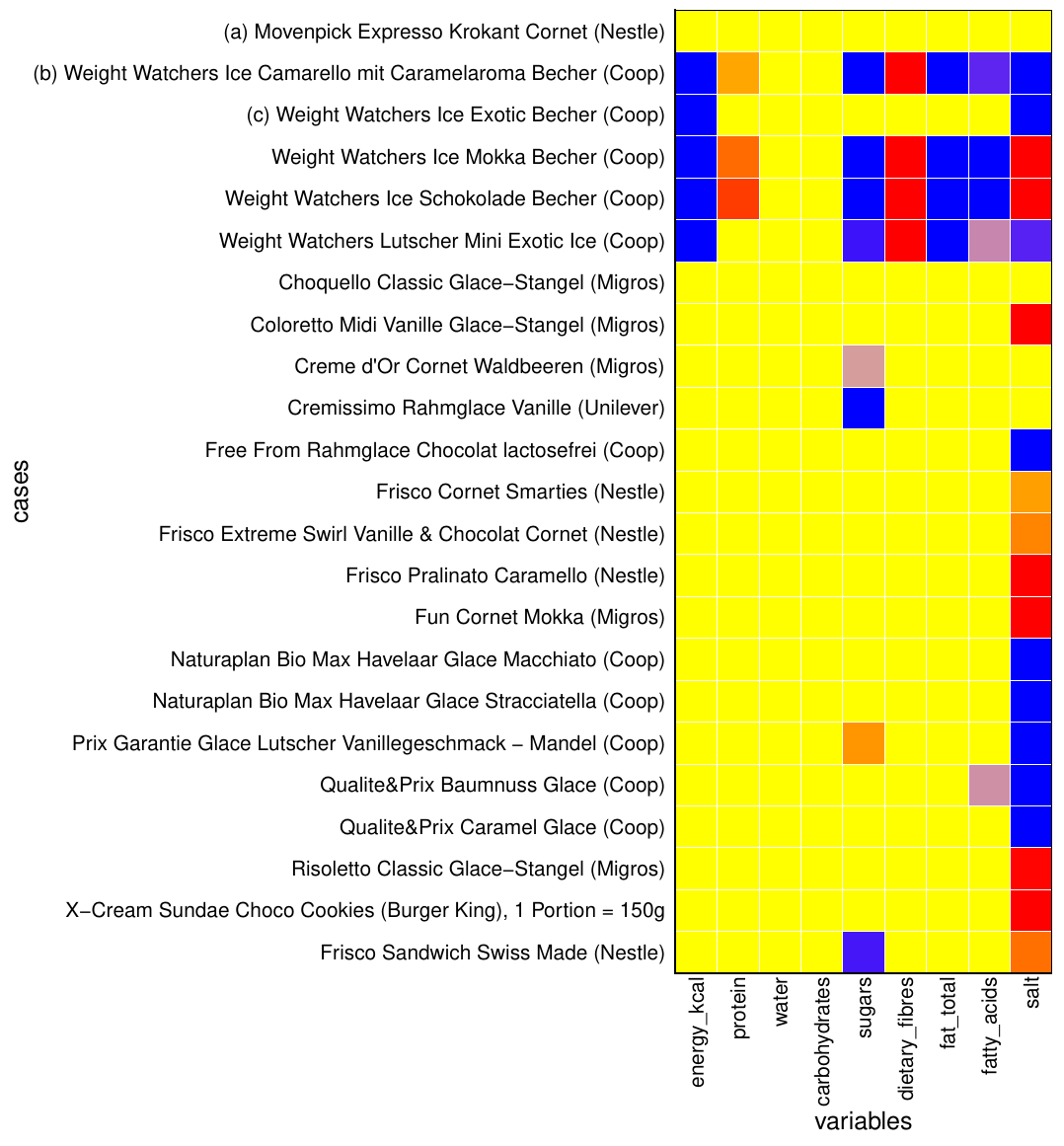}
\caption{Cellmap of selected ice cream products. 
Unflagged cells are yellow. Red cells have values 
that are substantially higher than expected, and 
blue cells lower than expected. The color 
intensity is proportional to the magnitude of the 
standardized residual.}
\label{fig:cellmap_selected}
\end{figure}

If we want to visualize more cellwise information
we naturally turn to a cellmap like
Figure~\ref{fig:cellmap_selected}. A cellmap 
contains a little square for each cell. The 
square is colored yellow for unflagged cells. It 
is red when the cell value is unexpectedly high, 
and blue when it is unexpectedly low. Here, high 
and low are relative to the predicted value of
the cell that was obtained by cellMCD in the 
given class. The cellmap in 
Figure~\ref{fig:cellmap_selected} contains some 
selected products from the ice cream class, 
including the points marked (a), (b) and
(c) in the classmap in 
Figure~\ref{fig:classmap_cellQDA}. Case (a) is
regular, and none of its cells are flagged
as they are all yellow.
The next five products in the list are from 
the Weight Watchers brand. As the name 
suggests, this brand positions itself 
in the market as a healthy option. Both RQDA and
cellQDA assign the Weight Watchers products to 
class 0. In Figure~\ref{fig:cellmap_selected} we 
see that cellQDA flags many cells of these 
products due to unexpectedly low energy, high
protein, low sugars, high dietary fibres, and
low fat. The remaining products in the cellmap
are those that were misclassified by RQDA but 
classified correctly by cellQDA. Looking at the 
cellmap, we attribute this mainly to the 
unexpectedly high and low salt levels.

\begin{figure}[!ht]
\centering
\includegraphics[width=0.6\linewidth]
    {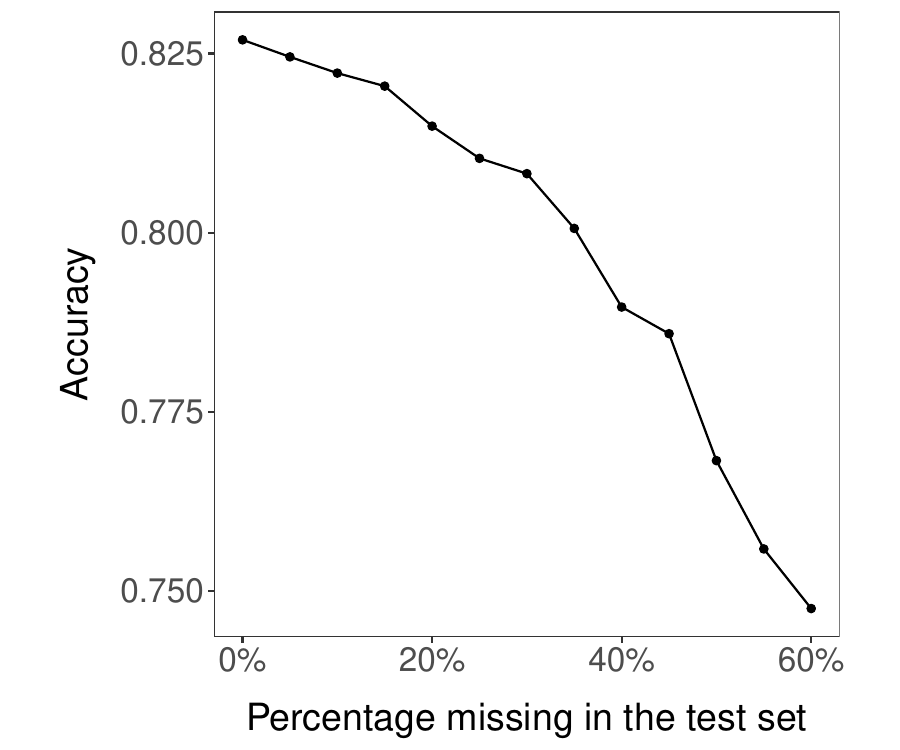}
\caption{Cross-validated accuracy of cellQDA 
on the sweets data, when an increasing 
percentage of random cells are set to 
missing in the test data.}
\label{fig:missing}
\end{figure}

Finally, we investigate how well cellQDA handles 
missing data by setting some random cells to missing 
in the test data. Figure \ref{fig:missing} shows 
the resulting cross-validated accuracy of cellQDA 
for up to 60\% of missing cells. For up to about 25\%
of NA's the accuracy decreases only slightly, and
afterward it goes down faster. But even at 60\% of
NA's, the 76\% accuracy of cellQDA still outperforms 
the 57\% accuracy of CQDA applied to complete 
test data.

Section~\ref{supp:sweets} of the Supplementary 
Material provides some more results on the sweets 
data, and Section~\ref{supp:other} looks into two 
other datasets.

%==================================================
\section{Conclusions}
\label{sec:conc}

We have introduced the cellwise robust discriminant 
analysis method cellQDA and its linear counterpart 
cellLDA. The main novelty of our approach lies in 
the out-of-sample prediction stage, which is able 
to classify cases in the test set that contain 
cellwise outliers themselves. In order to achieve
this we have proposed an extension of the cellwise
contamination model, as well as a likelihood-based
cellflagger for detecting outlying cells. The
approach is \mbox{coherent} in that the methods to
estimate
the class parameters, to detect outlying cells, and 
to assign test cases to classes, all use the same
underlying principle based on a penalized likelihood
function. This also enables it to handle missing
values in the data.

Extensive simulations indicate that cellQDA and 
cellLDA consistently outperform their classical 
counterparts as well as existing robust methods 
when cellwise outliers are present, without 
sacrificing performance when the data are clean 
or contain casewise outliers only. The methodology
is illustrated on a real life dataset about the
composition of food products, and the results 
are interpreted with the aid of graphical 
displays that further highlight the practical 
utility of this approach. 

Several paths for future research remain open. 
One interesting direction would be to extend 
this methodology to high-dimensional settings. 
Combining the cellwise robustness of our approach 
with regularized covariance estimation might
provide a powerful tool for large applications.
Additionally, investigating different 
distributional assumptions could further broaden 
the scope of the methodology.

\vspace{5mm}
\noindent{\bf Software availability.} A zip 
file with the R code, an example script, and 
datasets is at
\url{https://wis.kuleuven.be/statdatascience/code/cellda_r_code.zip}\,.

\small
\spacingset{1.2}

%=========== SUPPLEMENTARY MATERIAL ============

\clearpage
\pagenumbering{arabic}
% restarts page numbering from 1
% \setcounter{page}{1}
\appendix
%\section*{Supplementary Material}% \label{sec:A}
   % The * makes this section unnumbered
\begin{center}
\phantom{abc}\\
%\vspace{10mm}

\Large{Supplementary Material to:\\ 
  Cellwise Robust Discriminant Analysis}
\end{center}
\vspace{10mm}

\setcounter{equation}{0} % equation number restart at 1
\renewcommand{\theequation}
  {A.\arabic{equation}} % labels equations as (A.1),...

\spacingset{1.5} % changed for the SuppMat only

\section{Additional simulation results}
\label{supp:simul}

\subsection{Quadratic discriminant analysis}

Here we show the results of the counterpart of the 
simulation study in the main text, this time 
in $d=5$ dimensions. Everything else is set up
in the same way.

\begin{figure}[H]
\centering
\includegraphics[width=0.9\linewidth]
  {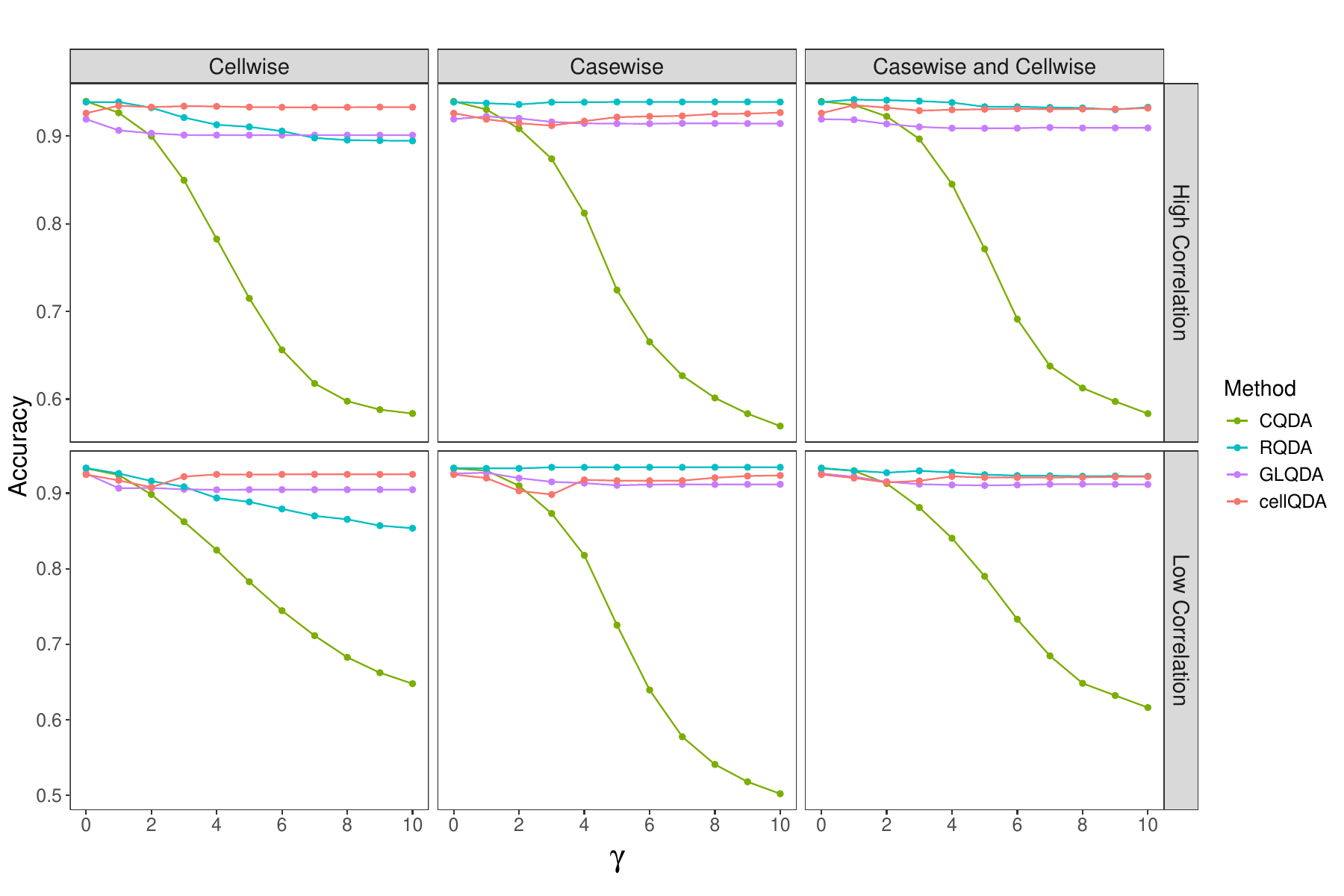}
\vspace{-2mm}
\caption{Accuracy of four QDA methods on 3
classes in 5 dimensions, with $n=200$. 
In the top row the classes have high within 
correlations, and the bottom row has low within 
correlations. The first column uses training
data with 10\% of cellwise outliers only, the 
second with 10\% of casewise outliers only, and
the third with 5\% of cellwise and 5\% of 
casewise outliers. The parameter $\gamma$ on the
horizontal axis says how far away the outliers 
are. The accuracy is measured on test data that
contains no contamination.}
\label{fig:QDA_clean_test_d5}
\end{figure}

\begin{figure}[H]
\centering
\includegraphics[width=0.9\linewidth]
  {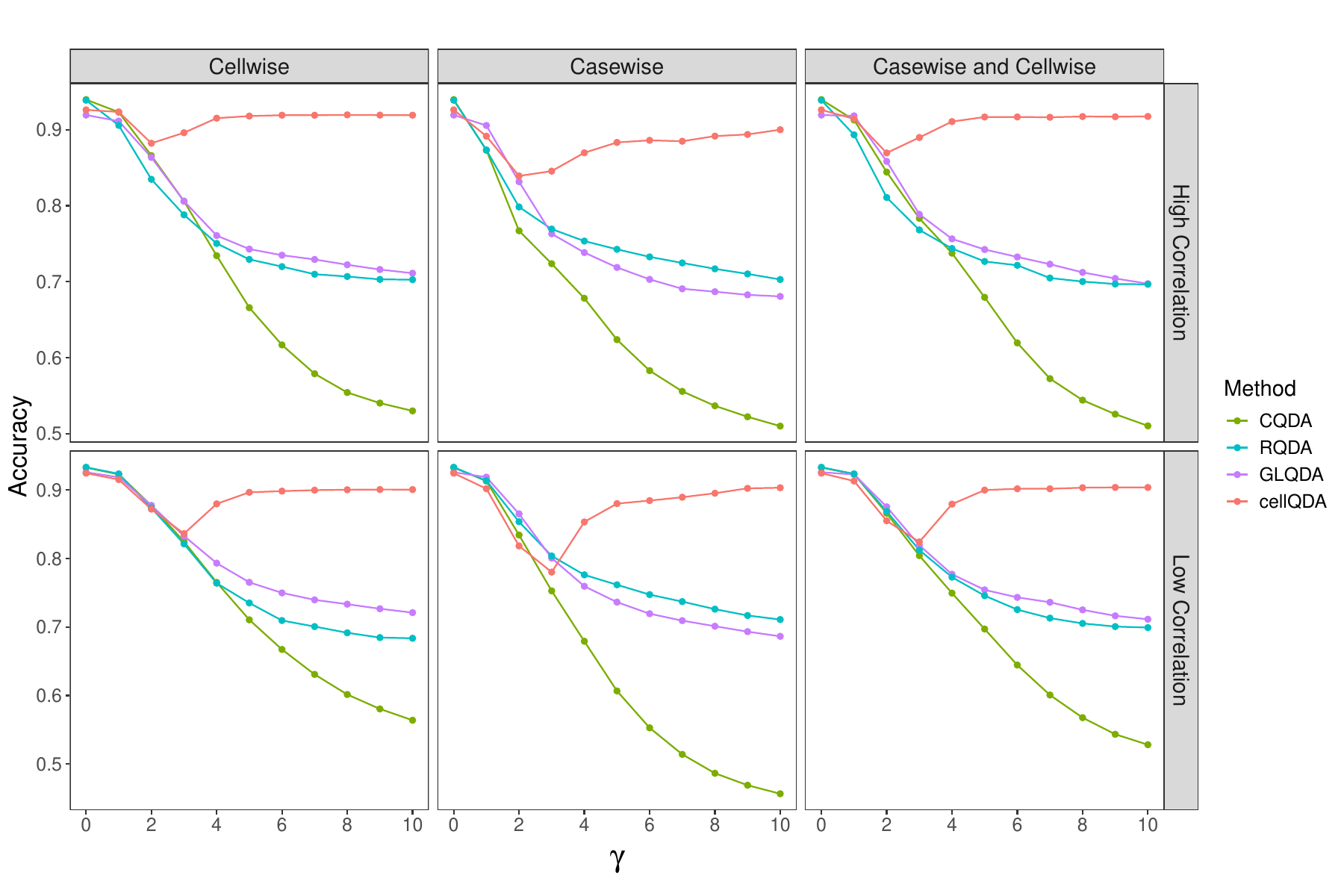}
\vspace{-4mm}
\caption{Same training data and fit as in 
  Figure~\ref{fig:QDA_clean_test_d5}, but now
  the test data has 10\% of outlying cells only.}
\label{fig:QDA_cell_test_d5}  
\end{figure}

\begin{figure}[H]
\vspace{-4mm}
\centering
\includegraphics[width=0.9\linewidth]
  {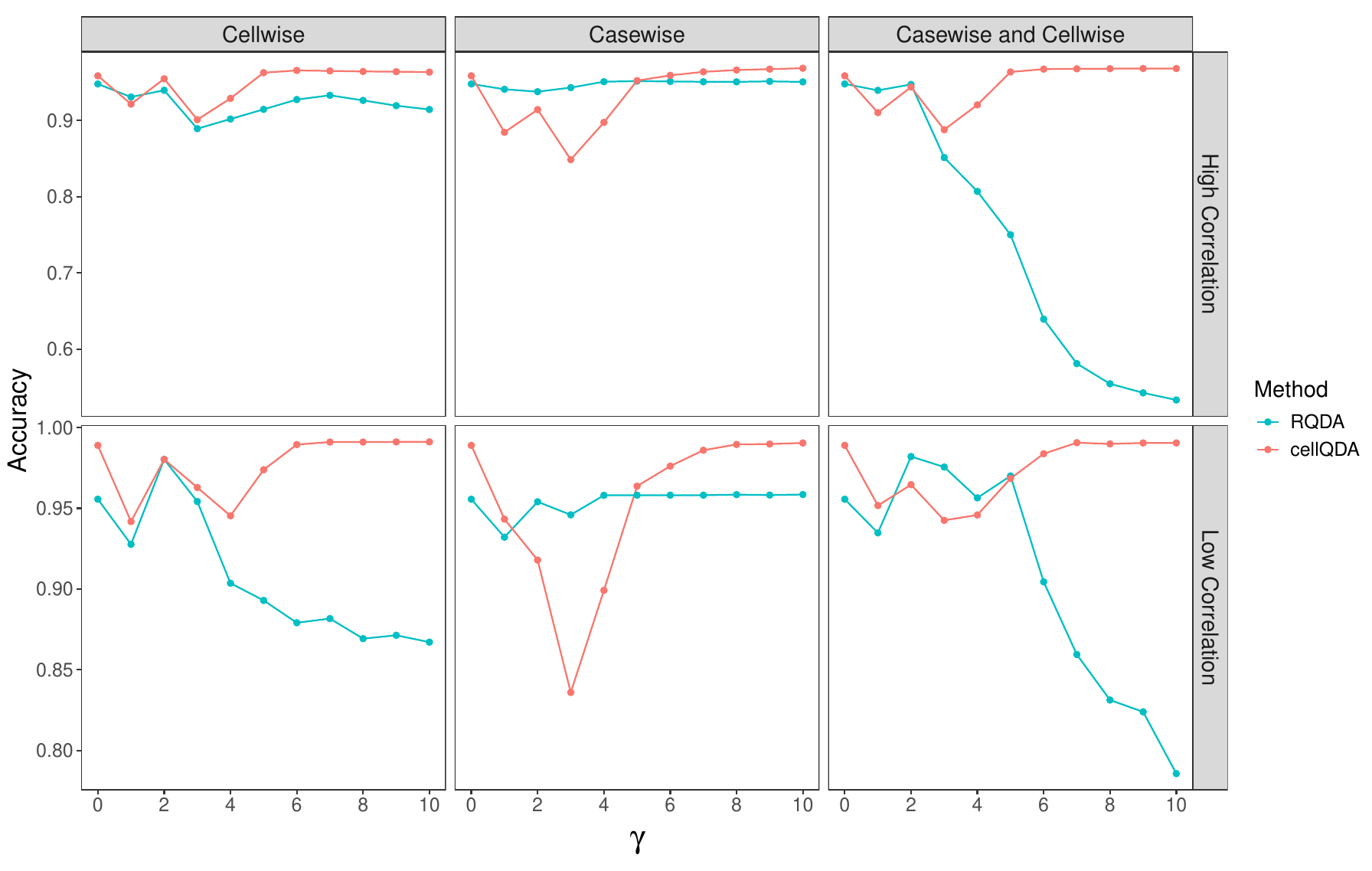}
\vspace{-4mm}
\caption{Comparison of RQDA and cellQDA on
the same training data as in 
Figure~\ref{fig:QDA_clean_test_d5} and displayed 
in the same way. Now the test data contains a 
class 0 with 10\% of generated casewise outliers. 
The reported accuracy is of the assignments 
to the four classes 0, 1, 2, and 3.}
\end{figure}

%===============================================
\subsection{Linear discriminant analysis}

In the main text, Figure~\ref{fig:LDA_d20}
showed the accuracy of the LDA methods when
the test data has 10\% of cellwise outliers
only. We will now look at the other kinds of
test data, also for $d=20$.

\begin{figure}[!ht]
\centering
\includegraphics[width=0.85\linewidth]
  {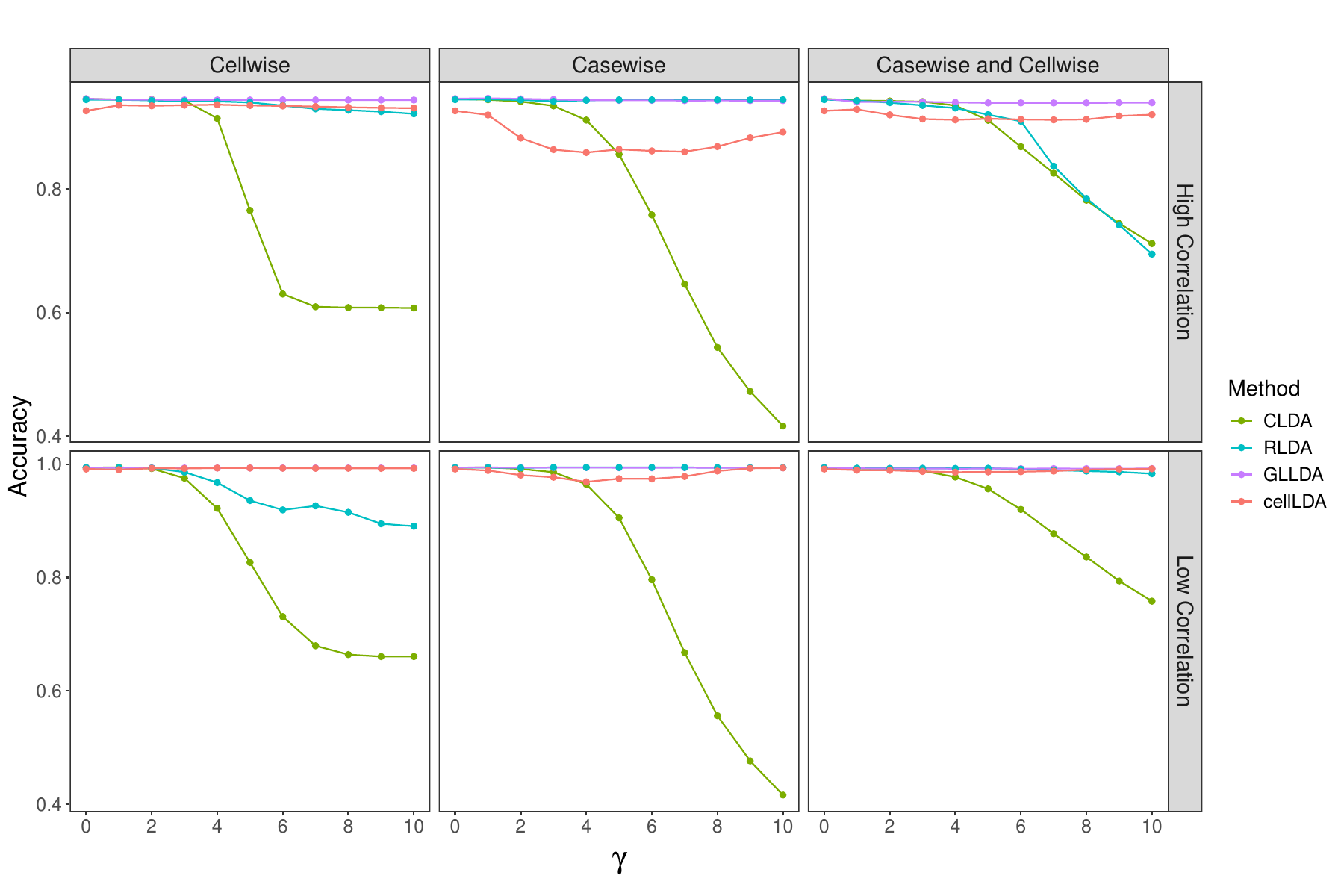}
\vspace{-4mm}
\caption{Like Figure~\ref{fig:LDA_d20} with
  the same training data, % and estimates, 
  but here the test data are clean.}
\label{fig:LDA_clean_test_d20}
\end{figure}

\begin{figure}[!ht]
\vspace{-4mm}
\centering
\includegraphics[width=0.85\linewidth]
  {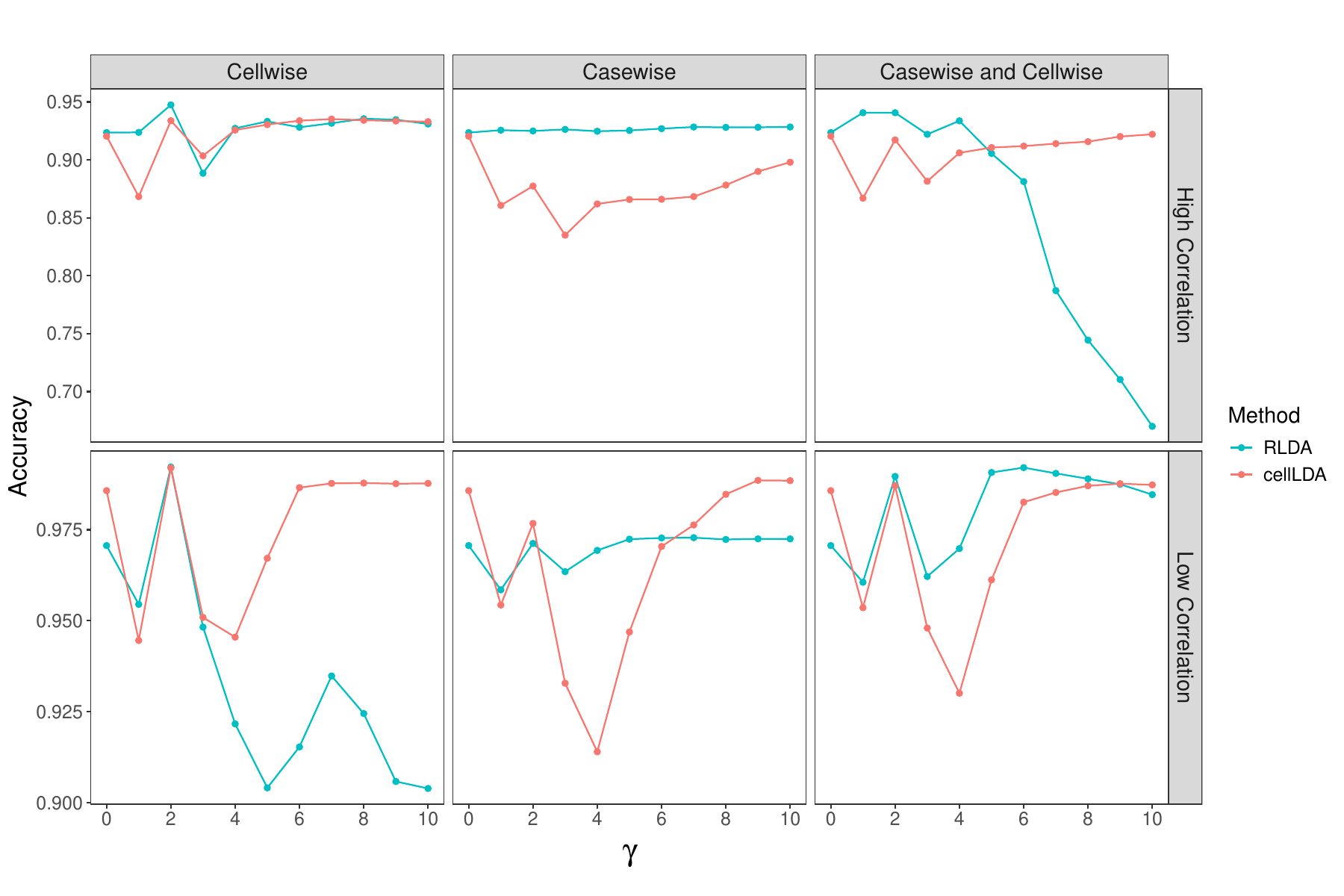}
\vspace{-4mm}
\caption{Like Figure~\ref{fig:LDA_d20} with
  the same training data and estimates, but
  now the test data has 10\% of outlying
  cases only.}
\label{fig:LDA_case_test_d20}
\end{figure}

We conclude that in $d=20$ dimensions, the
effects of outliers on the LDA methods are
similar to those on the corresponding QDA
methods. 

Below we see that the situation is similar for 
$d=5$ dimensions.

\begin{figure}[!ht]
\centering
\includegraphics[width=0.85\linewidth]
  {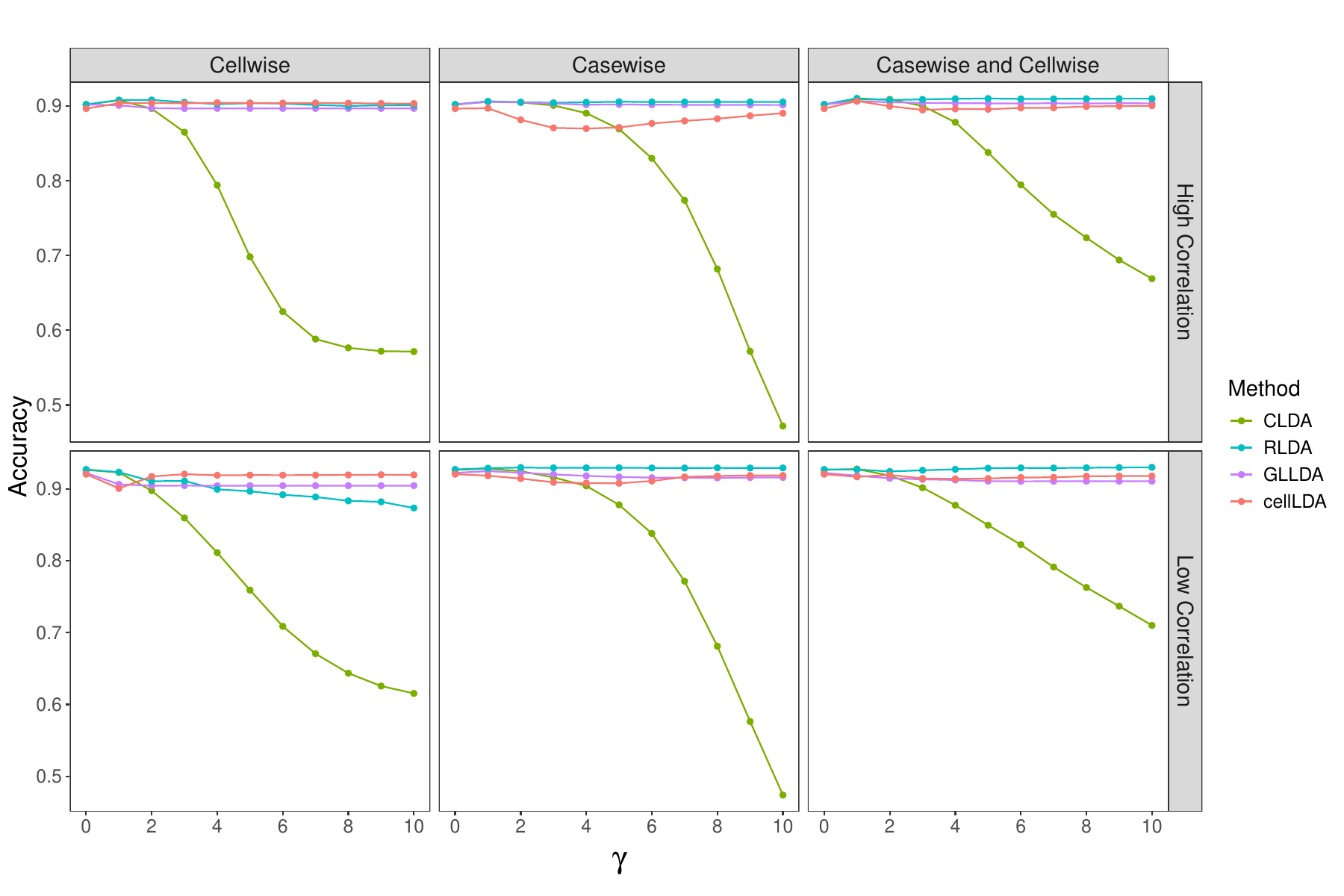}
\vspace{-3mm}
\caption{LDA. Like 
  Figure~\ref{fig:LDA_clean_test_d20} 
  with clean test data, but for $d=5$.}
\end{figure}

\begin{figure}[!ht]
\vspace{-2mm}
\centering
\includegraphics[width=0.85\linewidth]
  {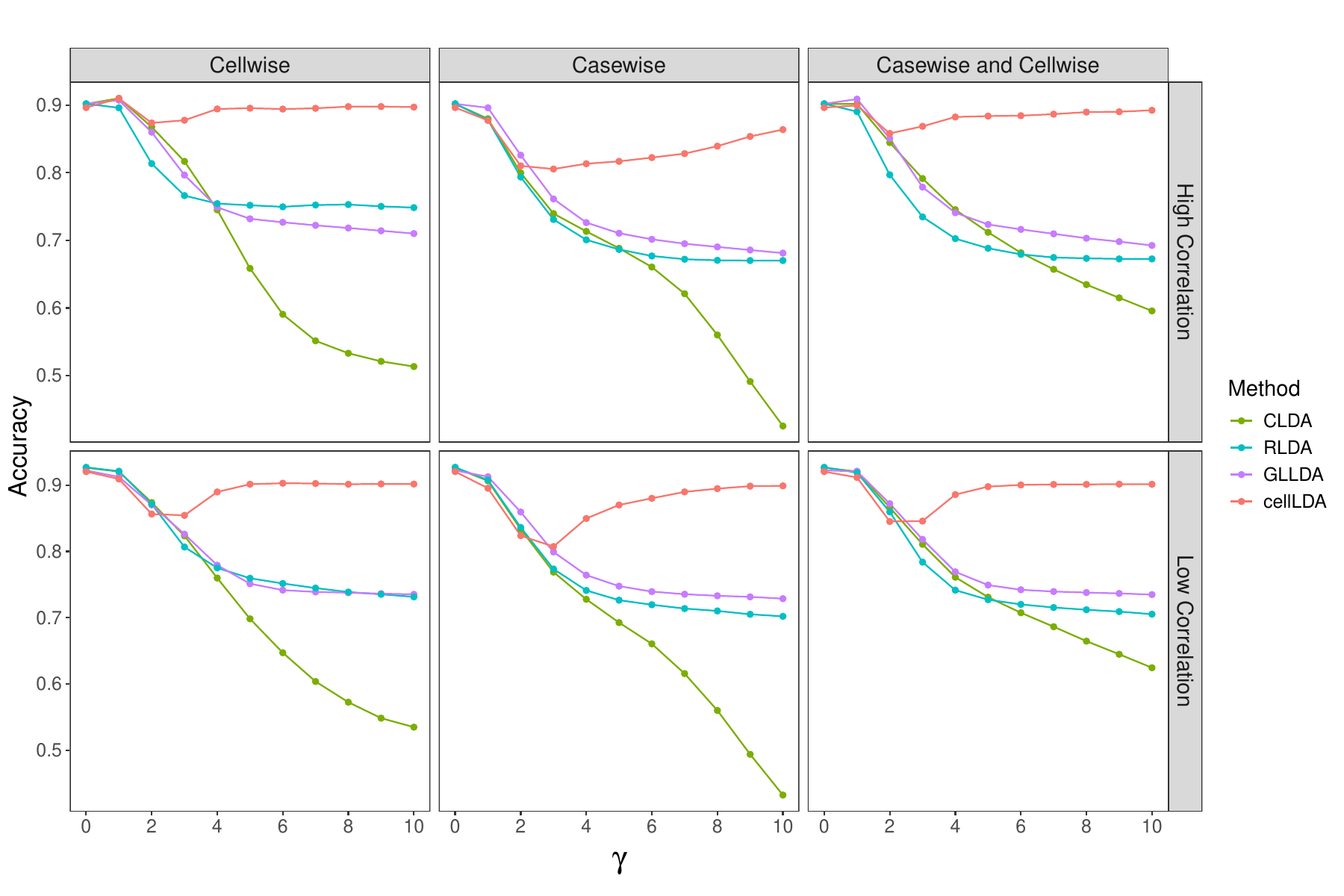}
\vspace{-3mm}
\caption{LDA. Like
  Figure~\ref{fig:LDA_d20} with test data
  containing cellwise outliers, but for $d=5$.}
\end{figure}

\begin{figure}[!ht]
\centering
\includegraphics[width=0.85\linewidth]
  {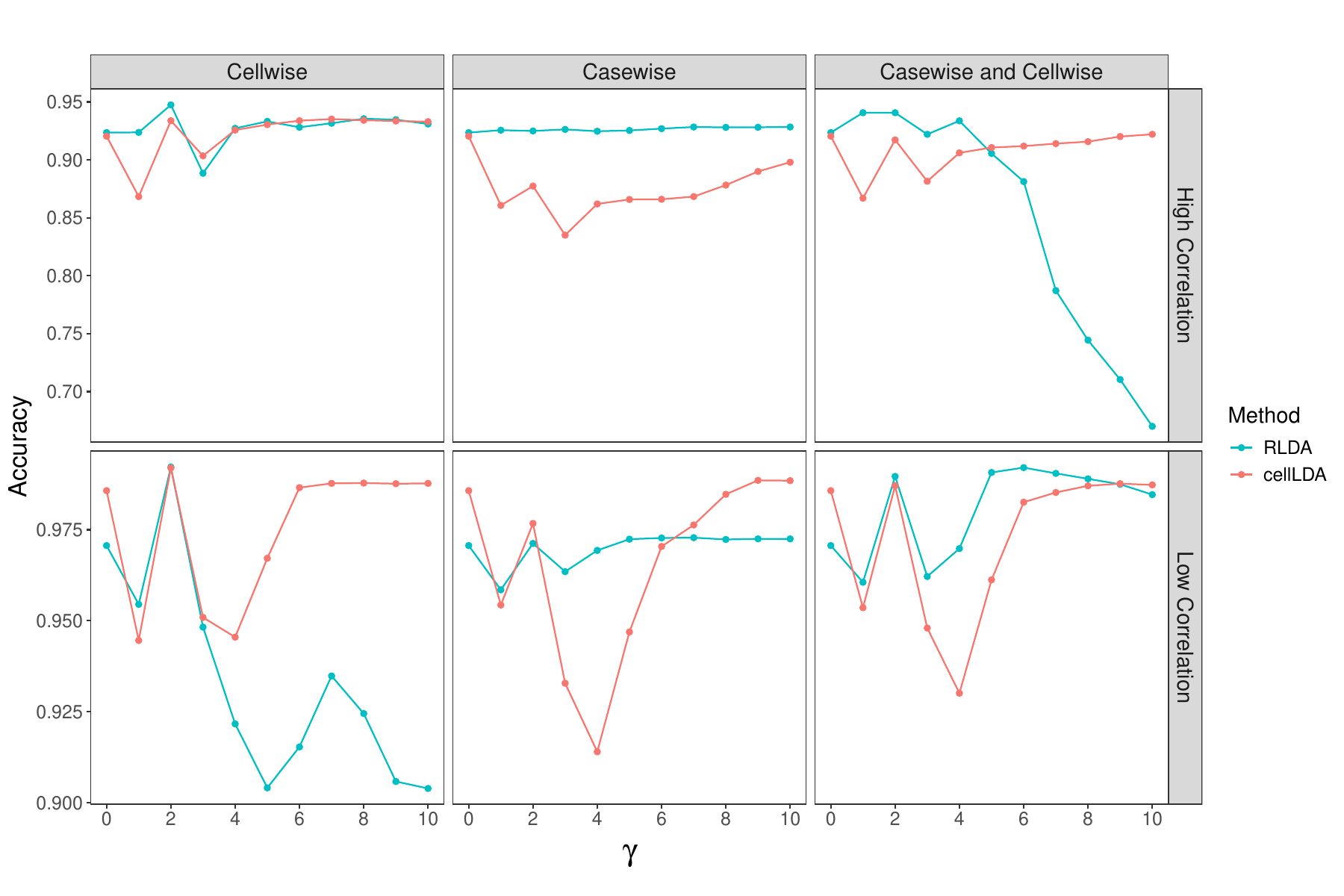}
\vspace{-3mm}    
\caption{LDA. Like
  Figure~\ref{fig:LDA_case_test_d20} with test 
  data containing casewise outliers, but for 
  $d=5$.} 
\end{figure}

Overall, the effects of cellwise and casewise 
outliers in 5 dimensions are similar to those 
in 20 dimensions, and those on LDA are similar
to those on QDA.

%==============================================
\clearpage
\section{Additional Results on the sweets data}
\label{supp:sweets}

In the main text we analyzed the sweets data
with cellQDA only. Here we also look at the
results of the methods that we compared with
in the simulation study. We randomly split off
a test dataset with 162 cases (that is, one 
fifth of the data), and trained on the
remainder. Table~\ref{tab:confusion_matrix} 
shows the confusion matrices on the test 
data.

\vspace{3mm}
\begin{table}[!ht]
\centering
\caption{Confusion matrices on test data. The 
rows contain the given classes, and the columns
contain the predictions. Here the class 0 
assignment of cellQDA and RQDA is turned off.}
\label{tab:confusion_matrix}
\begin{tabular}{llrrrr}
\toprule
 & & \multicolumn{4}{c}{\textit{Predicted class}} \\
\cmidrule(lr){3-6}
Method & \textit{True class} & biscuits & ice cream & cakes & puddings \\
\midrule
\multirow{4}{*}{CQDA} & biscuits & \textbf{30} & 4 & 19 & 0 \\
 & \cellcolor{gray!15} ice cream & \cellcolor{gray!15} 0 & \cellcolor{gray!15} \textbf{42} & \cellcolor{gray!15} 2 & \cellcolor{gray!15} 0 \\
 & cakes & 0 & 9 & \textbf{33} & 0 \\
 & \cellcolor{gray!15} puddings & \cellcolor{gray!15} 2 & \cellcolor{gray!15} 20 & \cellcolor{gray!15} 0 & \cellcolor{gray!15} \textbf{1} \\
\midrule
\multirow{4}{*}{RQDA} & biscuits & \textbf{47} & 0 & 6 & 0 \\
 & \cellcolor{gray!15} ice cream & \cellcolor{gray!15} 0 & \cellcolor{gray!15} \textbf{36} & \cellcolor{gray!15} 6 & \cellcolor{gray!15} 2 \\
 & cakes & 2 & 0 & \textbf{40} & 0 \\
 & \cellcolor{gray!15} puddings & \cellcolor{gray!15} 2 & \cellcolor{gray!15} 2 & \cellcolor{gray!15} 5 & \cellcolor{gray!15} \textbf{14} \\
\midrule
\multirow{4}{*}{GLQDA} & biscuits & \textbf{48} & 0 & 5 & 0 \\
 & \cellcolor{gray!15} ice cream & \cellcolor{gray!15} 0 & \cellcolor{gray!15} \textbf{42} & \cellcolor{gray!15} 2 & \cellcolor{gray!15} 0 \\
 & cakes & 4 & 7 & \textbf{31} & 0 \\
 & \cellcolor{gray!15} puddings & \cellcolor{gray!15} 2 & \cellcolor{gray!15} 19 & \cellcolor{gray!15} 2 & \cellcolor{gray!15} \textbf{0} \\
\midrule
\multirow{4}{*}{cellQDA} & biscuits & \textbf{47} & 0 & 6 & 0 \\
 & \cellcolor{gray!15} ice cream & \cellcolor{gray!15} 0 & \cellcolor{gray!15} \textbf{35} & \cellcolor{gray!15} 4 & \cellcolor{gray!15} 5 \\
 & cakes & 2 & 0 & \textbf{40} & 0 \\
 & \cellcolor{gray!15} puddings & \cellcolor{gray!15} 2 & \cellcolor{gray!15} 2 & \cellcolor{gray!15} 4 & \cellcolor{gray!15} \textbf{15} \\
\bottomrule
\end{tabular}
\end{table} 

Note that CQDA almost never assigns a product to the 
puddings class, instead most puddings are classified as 
ice cream. It also classifies many biscuits as cakes.
GLQDA did not assign any product to the puddings
class. In contrast, the confusion matrices of RQDA 
and cellQDA are dominated by their diagonals, meaning
that most products were assigned to the correct
class.

\begin{figure}[H] 
\centering
\includegraphics[width=0.95\linewidth]
   {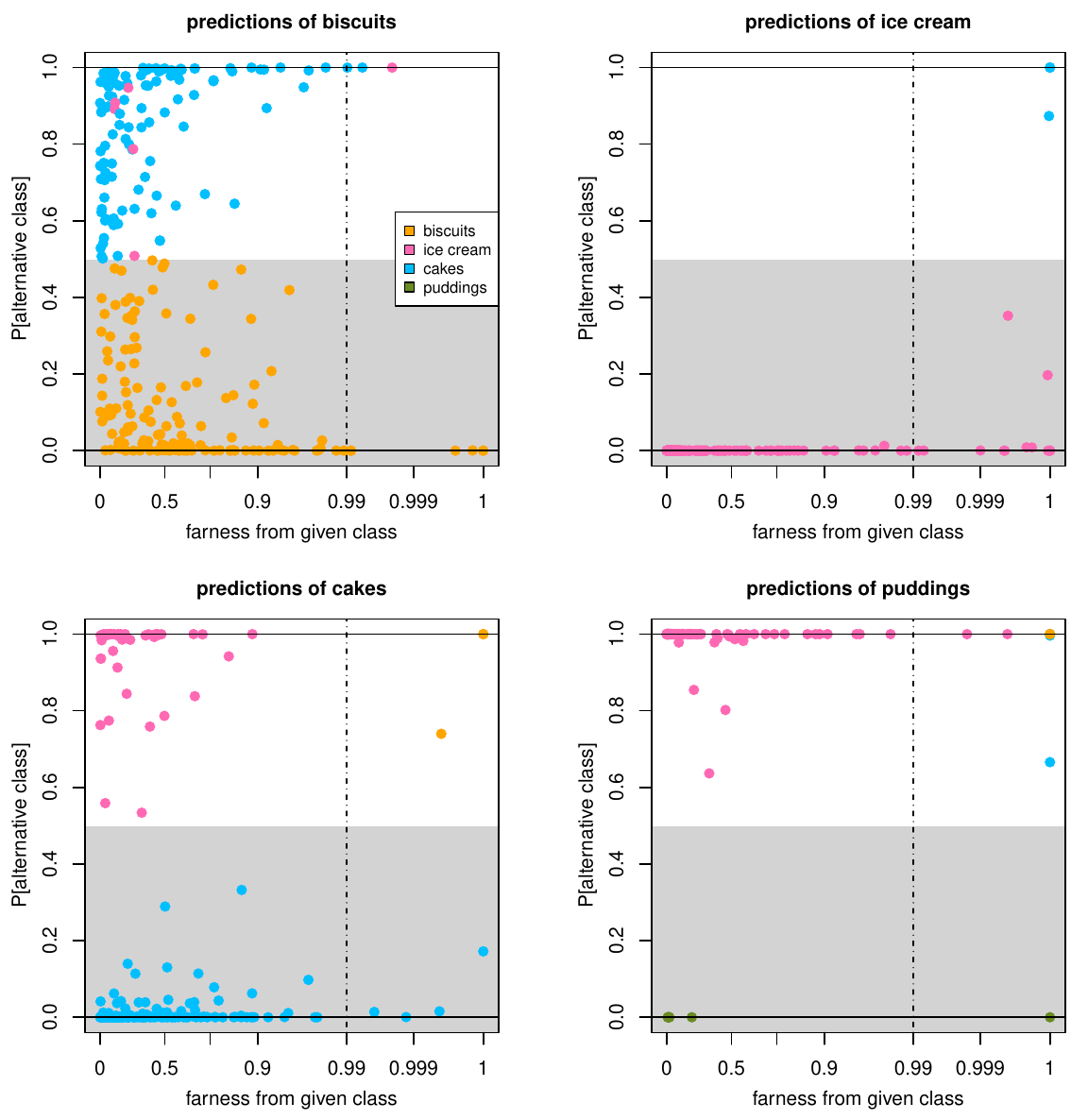}
\caption{Classmap of CQDA on the sweets data.}
\label{fig:classmap_CQDA}
\end{figure}

\begin{figure}[H] 
\centering
\includegraphics[width=0.95\linewidth]
  {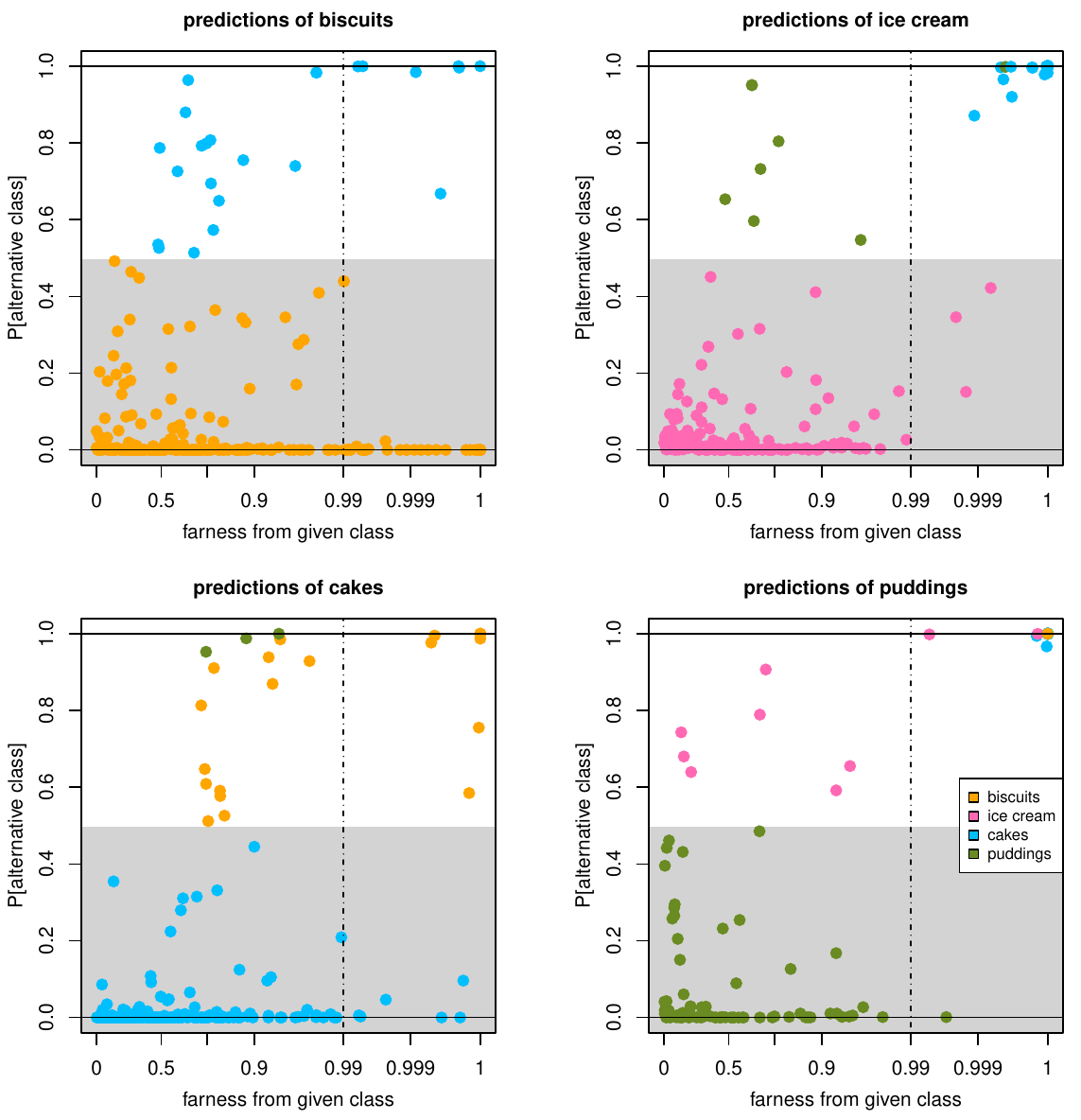}
\caption{Classmap of RQDA on the sweets data.}
\label{fig:classmap_RQDA}
\end{figure}

\begin{figure}[H] 
\centering
\includegraphics[width=0.95\linewidth]
  {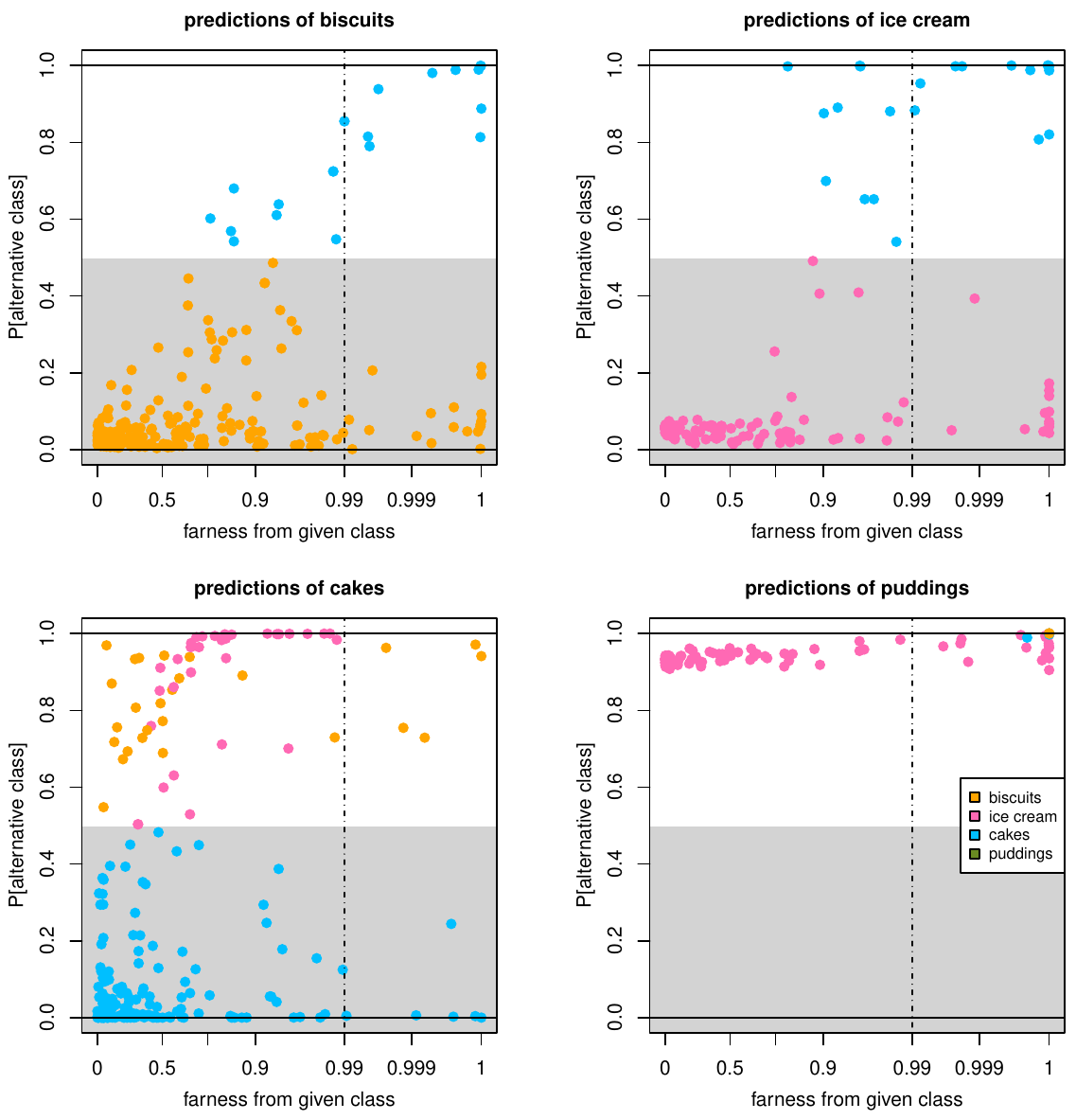}
\caption{Classmap of GLQDA on the sweets data.}
\label{fig:classmap_GLQDA}
\end{figure}

The same effects are visible in more detail in the
classmaps of these methods, shown in Figures 
\ref{fig:classmap_CQDA}-\ref{fig:classmap_GLQDA}.\\

%============================================
\section{Two other examples}
\label{supp:other}

%============================================
\subsection{Phoneme dataset}

The phoneme dataset contains digitized speech 
signals used to distinguish between nasal and 
oral phonemes. It has been analyzed in the 
literature on regularized discriminant analysis,
including the work of \citet{aerts2017}. Our
analysis uses $d=20$ features that correspond 
to 20 different log-periodogram frequencies, 
and $n=1000$ cases.
The original dataset is publicly available on
\url{http://www-stat.stanford.edu/ElemStatLearn}\,.
The subset used here  can be downloaded from 
\url{https://wis.kuleuven.be/statdatascience/code/cellda_r_code.zip}\,.

We conduct two experiments. First, we compare all 
four methods on the original data to assess their 
baseline classification performance. There the 
5-fold cross-validation accuracy of cellQDA is 
0.815, and exceeds those of CQDA (0.762), 
RQDA (0.764), and GLQDA (0.772) that lie close 
together.

Second, we artificially contaminate the data with
both cellwise and casewise outliers as in the 
simulation design in Section~\ref{sec:sim}, but
now based on the estimated $\bhmu_g$ and 
$\bhSigma_g$ as we have no true values.
We generate 5\% of cellwise outliers and 5\% of
casewise outliers. The accuracy was obtained from
$5$-fold cross-validation, that was repeated 
$10$ times. The casewise outliers were excluded 
from the test folds, in order to enable a fair 
comparison across methods with different outlier 
detection capabilities.

\begin{figure}[H]
\centering
\includegraphics[width=0.7\linewidth]
   {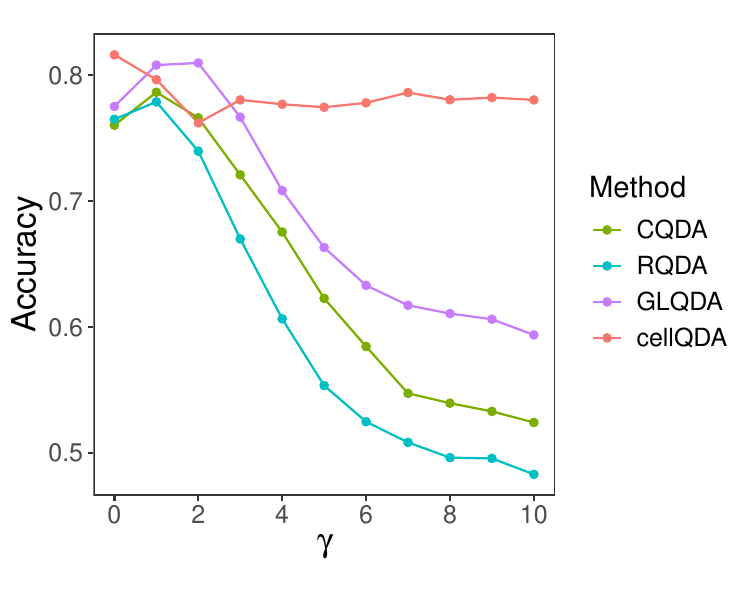}
\vspace{-7mm}
\caption{Cross-validated accuracy on the phoneme 
data. For $\gamma = 0$ we used the original data.
For $\gamma>0$, we inserted 5\% of cellwise and 
5\% of casewise outliers.}
\label{fig:phoneme_accuracy}
\end{figure}

In Figure~\ref{fig:phoneme_accuracy} we see that
contaminating the data does not affect cellQDA 
much, but the other methods suffer from it, 
especially as $\gamma$ increases.

%==============================================
\subsection{Yeast dataset}

The yeast dataset is a widely used benchmark for 
protein localization prediction \citep{yeast1991}. 
The original data 
contain 10 classes, corresponding to different 
cellular localizations, but several classes have 
very few cases. We restrict our analysis to the 
three largest classes: CYT (cytosolic), 
NUC (nuclear), and MIT (mitochondrial), yielding 
a total of $n = 665$ cases.
We also remove three variables from the analysis.
The variables \texttt{erl} and \texttt{pox} are 
excluded because they are nearly constant across
all cases. The variable \texttt{nuc} exhibits 
irregular behavior: it is discrete or even constant 
in some classes rather than continuous, so it is 
excluded too. After these preprocessing steps, 
$d = 5$ variables remain. 
The original dataset is publicly available 
through the UCI Machine Learning Repository at 
\url{https://doi.org/10.24432/C5KG68}\,.
The subset can be downloaded from 
\url{https://wis.kuleuven.be/statdatascience/code/cellda_r_code.zip}\,.

\begin{figure}[H]
\centering
\includegraphics[width=0.7\linewidth] 
  {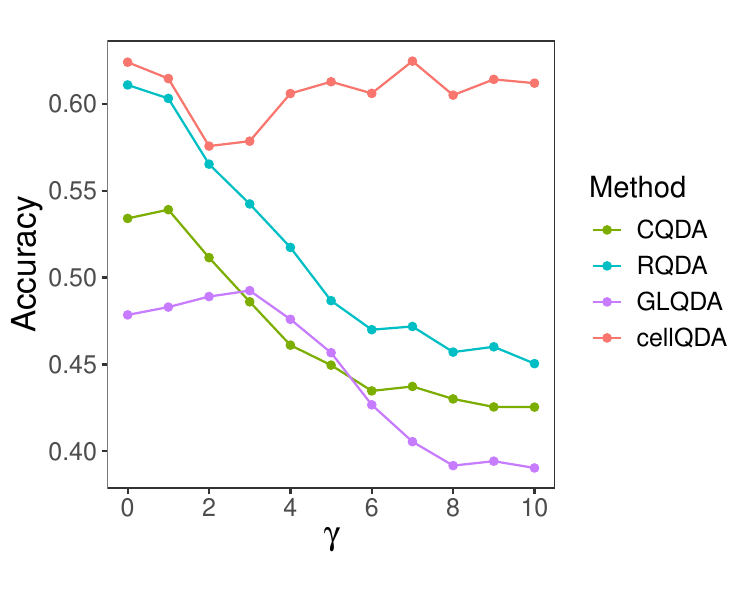}
\vspace{-7mm}  
\caption{Cross-validated accuracy on the yeast 
data. For $\gamma = 0$ we used the original data. 
For $\gamma>0$, we inserted 5\% of cellwise and 
5\% of casewise outliers.}
\label{fig:yeast_accuracy}
\end{figure}

We carry out the same experiments as on the phoneme 
dataset. On the original data, the accuracy of cellQDA 
is 0.627, and exceeds those of CQDA (0.532), RQDA 
(0.614), and GLQDA (0.478). All accuracies are fairly 
low on these data, because the classes CYT and NUC
overlap substantially. 
Figure~\ref{fig:yeast_accuracy} indicates that the
accuracy of cellQDA is not affected much by the 
added contamination.

\end{document}